\documentclass[journal]{IEEEtran}
\usepackage{graphicx}
\usepackage{caption}
\usepackage{enumitem}
\usepackage{multicol}
\usepackage{amsmath,amssymb}
\usepackage{amsfonts}
\usepackage{textcomp}
\usepackage{psfrag}
\usepackage{multimedia}
\usepackage{fancybox}

\newcommand{\vect}[1]{\ensuremath{\boldsymbol{\mathrm{#1}}}}

\newtheorem{Proposition}{Proposition}

\newtheorem{Assumption}{Assumption}

\usepackage[usenames,dvipsnames]{color}
\definecolor{wheat}{rgb}{0.96,0.87,0.70}
\definecolor{mario}{rgb}{0.8,0.8,1}
\definecolor{seb}{rgb}{0.8,1,0.8}

\newcommand {\matr}[2]{\left[\begin{array}{#1}#2\end{array}\right]}

\newcounter{lastnote}

\usepackage[ruled,vlined]{algorithm2e}

\begin{document} 

\title{Towards Safe Reinforcement Learning Using NMPC and Policy Gradients: Part II - Deterministic Case} 

\author
{S\'ebastien Gros, Mario Zanon
	\thanks{S\'ebastien Gros is with the Department of Cybernetic, NTNU, Norway.}
	\thanks{Mario Zanon is with the IMT School for Advanced Studies Lucca, Lucca
		55100, Italy.}
}



%
%

\IEEEtitleabstractindextext{
	\begin{abstract} In this paper, we present a methodology to deploy the deterministic policy gradient method, using actor-critic techniques, when the optimal policy is approximated using a parametric optimization problem, where safety is enforced via hard constraints. For continuous input space, imposing safety restrictions on the exploration needed to deploying the deterministic policy gradient method poses some technical difficulties, which we address here. We will investigate in particular policy approximations based on robust Nonlinear Model Predictive Control (NMPC), where safety can be treated explicitly. For the sake of brevity, we will detail the construction of the safe scheme in the robust linear MPC context only. The extension to the nonlinear case is possible but more complex. We will additionally present a technique to maintain the system safety throughout the learning process in the context of robust linear MPC. This paper has a companion paper treating the stochastic policy gradient case.
	\end{abstract}
	
}

\maketitle

\IEEEdisplaynontitleabstractindextext

\IEEEpeerreviewmaketitle

\section{Introduction}
Reinforcement Learning (RL) is a powerful tool for tackling Markov Decision Processes (MDP) without depending on a detailed model of the probability distributions underlying the state transitions. Indeed, most RL methods rely purely on observed state transitions, and realizations of the stage cost $L(\vect s,\vect a)\in\mathbb{R}$ assigning a performance to each state-input pair $\vect s,\vect a$ (the inputs are often labelled actions in the RL community). RL methods seek to increase the closed-loop performance of the control policy deployed on the MDP as observations are collected. RL has drawn an increasingly large attention thanks to its accomplishments, such as, e.g., making it possible for robots to learn to walk or fly without supervision \cite{Wang2012b,Abbeel2007}. 

Most RL methods are based on learning the optimal control policy for the real system either directly, or indirectly. Indirect methods typically rely on learning a good approximation of the optimal action-value function underlying the MDP. The optimal policy is then indirectly obtained as the minimizer of the value-function approximation over the inputs $\vect a$. Direct RL methods, based on policy gradients, seek to adjust the parameters $\vect \theta$ of a given policy $\vect\pi_{\vect\theta}$ such that it yields the best closed-loop performance when deployed on the real system. An attractive advantage of direct RL methods over indirect ones is that they are based on formal necessary conditions of optimality for the closed-loop performance of $\vect\pi_{\vect\theta}$, and therefore asymptotically (for a large enough data set) guarantee the (possibly local) optimality of the parameters $\vect\theta$ \cite{Sutton1999, Silver2014}.

RL methods often rely on Deep Neural Networks (DNN) to carry the policy approximation $\vect\pi_{\vect\theta}$. While effective in practice, control policies based on DNNs provide limited opportunities for formal verifications of the resulting closed-loop behavior, and for imposing hard constraints on the evolution of the state of the real system. The development of safe RL methods, which aims at tackling this issue, is currently an open field or research \cite{Garcia2015}. 

In this paper, we investigate the use of constrained parametric optimization problems to carry the policy approximation. The aim is to impose safety by means of hard constraints in the optimization problem. Most RL methods require exploration, i.e., the inputs applied to the real system must differ from the policy $\vect\pi_{\vect\theta}$ in order to identify changes in the policy parameters ${\vect\theta}$  that can yield a higher closed-loop performance. Exploration is typically performed via stochastic disturbances of the policy. We will show in this paper that the presence of hard constraints distorts the statistics of the exploration, and that some corrections must in theory be introduced in the classic tools underlying the deterministic policy gradient method to account for this distortion. We propose computationally efficient tools to implement these corrections, based on parametric Nonlinear Programming techniques, and interior-point methods.

Robust Nonlinear Model Predictive Control (NMPC) is arguably an ideal candidate for forming the constrained optimization problem supporting the policy approximation. 
Robust NMPC techniques provide safety guarantees on the closed-loop behavior of the system by explicitly accounting for the presence of (possibly stochastic) disturbances and model inaccuracies. A rich theoretical framework is available on the topic \cite{Mayne2014}. The policy parameters $\vect \theta$ will then appear as parameters in the NMPC model(s), cost function and constraints. Updates in the policy parameters $\vect \theta$ will then be driven by the deterministic policy gradient method to increase the NMPC closed-loop performance, and constrained by the requirement that the NMPC model inaccuracies must be adequately accounted for in forming the robust NMPC scheme. For the sake of brevity and simplicity, we will detail these questions in the specific linear MPC case. The extension to the nonlinear case is arguably possible, but more complex.

This paper has a companion paper \cite{Gros2019a} treating the same problem in the context of the stochastic policy gradient approach. The two papers share the same background material, and some similar techniques. However, the theory allowing the deployment of the two policy gradient techniques is intrinsically different.

The paper is structured as follows. Section \ref{sec:Background} provides some background material. Section \ref{sec:NMPCIntro} details the safe policy approximation we propose to use. Section \ref{sec:SafeExplo} establishes the basic properties that a safe exploration must fulfil in order to be able to build a correct policy gradient estimation with standard RL tools. Section \ref{eq:Numerics} presents an optimization-based approach to generate an exploration satisfying these properties. Section \ref{sec:RLsteps} discusses a technique to enforce safety in the RL-based learning process in the context of robust MPC. Section \ref{sec:Simulations} proposes an example of simulation using the principles developed in this paper.



\section{Background on Markov Decision Processes} \label{sec:Background}
This section provides background material on Markov Decision Processes (MDP), and on their restriction to a safe set. We also provide a brief introduction to the deterministic policy gradient method.

\subsection{Markov Decision Processes}
In the following, we will consider that the dynamics of the real system are described as a Markov Process (MP), with state transitions having the underlying conditional probability density:
\begin{align}
\label{eq:State:Transition}
\mathbb{P}\left[\vect{s}_{+}\,|\,\vect{s},\vect{a}\right]
\end{align}
denoting the probability density of the state transition $\vect{s},\vect{a}\to \vect{s}_{+}$. We will furthermore consider deterministic policies:
\begin{align}
\label{eq:DetPolicyDef}
\vect a = \vect{\pi}\left(\vect{s}\right)
\end{align}
associating an input (a.k.a. action) $\vect a \in \mathbb{R}^{n_{\vect a}}$ to any feasible state $\vect s \in \mathbb{R}^{n_{\vect s}}$. In the following, it will be additionally useful to introduce the concept of stochastic policy 
\begin{align}
\label{eq:StochPolicyDef}
{\pi}\left[\vect{a}\,|\,\vect{s}\right]\,:\, \mathbb{R}^{n_{\vect a}}\times  \mathbb{R}^{n_{\vect s}}\,\rightarrow \mathbb{R}_+
\end{align}
denoting the probability density of selecting a given input $\vect a$ for a given state $\vect s$. It is useful to observe that any deterministic policy \eqref{eq:DetPolicyDef} can be defined as a stochastic policy using:
\begin{align}
\label{eq:Stoch:Det:Equivalence}
{\pi}\left[\vect{a}\,|\,\vect{s}\right] = \delta\left(\vect{a} - \vect{\pi}\left(\vect{s}\right)\right)
\end{align}
where $\delta$ is the Dirac function. All the definitions below then readily apply to both \eqref{eq:StochPolicyDef} and \eqref{eq:DetPolicyDef} by using \eqref{eq:Stoch:Det:Equivalence}. Let us consider the distribution of the MP resulting from the state transition \eqref{eq:State:Transition} and policy \eqref{eq:StochPolicyDef}:
\begin{align}
\mathbb{P}\left[\vect{s}_k\,|\,{\pi}\right] = \int \prod_{i=0}^{k-1} &\mathbb{P}\left[\vect{s}_{i+1}\,|\,\vect{s}_i,\vect{a}\right] \mathbb{P}\left[\vect{s}_0\right]{\pi}\left[\vect{a}_i\,|\,\vect{s}_i\right]\\ &\mathrm{d}\vect{s}_{0,\ldots,k-1}\mathrm{d}\vect{a}_{0,\ldots,k-1}\nonumber
\end{align}
where $\mathbb{P}\left[\vect{s}_0\right]$ denotes the probability distribution of the initial conditions $\vect s_0$ of the MP. We can then define the discounted expected value of the MP distribution under policy ${\pi}$, labelled $\mathbb{E}_{{{\pi}}}[.]$, which reads as:
\begin{align}
\mathbb{E}_{{{\pi}}}[\zeta] := \sum_{k=0}^\infty \int \gamma^k& \zeta(\vect{s}_k, \vect{a}_k)  \mathbb{P}\left[\vect{s}_k\,|\,{\pi}\right]{\pi}\left[\vect{a}_k\,|\,\vect{s}_k\right]\mathrm{d}\vect{s}_k\mathrm{d}\vect{a}_k
\end{align}
\vspace{-0.5cm}\\
for any function $\zeta$. In the following we will assume the local stability of the MP under the selected policies. More specifically, we assume that $\pi$ is such that:
\begin{align}
\label{ass:Stability}
\lim_{\tilde{ \pi}\rightarrow{\pi}} \mathbb{E}_{{\tilde{ \pi}}}\left[\zeta\right] = \mathbb{E}_{{{\pi}}}\left[\zeta\right], 
\end{align}
for any function $\zeta$ such that both sides of the equality are finite. Assumption \eqref{ass:Stability} is underlying standard RL algorithms, though it is often left implicit, and allows us to draw equivalences between a policy and disturbances of that policy, which is required in the context of policy gradient methods. It can be construed as a local regularity assumption on $\mathbb{E}_{{{\pi}}}\left[.\right]$ that, e.g., holds if the system dynamics in closed-loop with policy $\pi$ are stable.

For a given stage cost function $L(\vect s,\vect a)$ and a discount factor $\gamma \in [0,1]$, the performance of a policy $\pi$ is given by the discounted cost:
\begin{align}
\label{eq:Return}
J(\pi) = \mathbb{E}_{{\pi}}\left[\,  L\, \right] 
\end{align}
The state transition \eqref{eq:State:Transition}, stage cost $L$ and discount factor $\gamma$ define a Markov Decision Process, with an underlying optimal policy given by:
\begin{align}
\label{eq:OptimalPolicy}
\pi_\star =\mathrm{arg} \min_{\pi}\, J(\pi)
\end{align}
It is useful to underline here that, while \eqref{eq:OptimalPolicy} may have several (global) solutions, any fully observable MDP admits a deterministic policy $\vect\pi_\star$ among its solutions.

The (scalar) value function and action-value functions associated to a given policy $\pi$ are given by \cite{Bertsekas1995,Bertsekas1996,Bertsekas2007}:
\begin{subequations}
\label{eq:Bellman:Policy:0}
\begin{align}
Q_{\pi}\left(\vect s,\vect a\right) &= L(\vect s,\vect a) + \gamma \mathbb{E}\left[V_{\pi}(\vect s_{+})\,|\, \vect s,\, \vect a\right],  \label{eq:MDP:Qfunction:Generic}\\
V_{\pi}\left(\vect s\right) &= \mathbb{E}_{\vect a\sim \pi[\cdot|\vect x]}\left[Q_{\pi}\left(\vect s, \vect a\right)\right], 
\end{align}
\end{subequations}
where the expected value in \eqref{eq:MDP:Qfunction:Generic} is taken over state transitions \eqref{eq:State:Transition}. The advantage function is then defined as:
\begin{align}
\label{eq:A:definition}
A_{\pi}\left(\vect s,\vect a\right) = Q_{\pi}\left(\vect s,\vect a\right) - V_{\pi}\left(\vect s\right) 
\end{align}
and provides the value of using input $\vect a$ in a given state $\vect s$ compared to using the policy $\pi$. Furthermore
\begin{align}
\label{eq:SomeAProperties}
A_{\vect\pi_\star}\left(\vect s,\vect a\right) \geq 0,\quad \forall\, \vect s,\vect a
\end{align}
holds at the deterministic optimal policy $\vect\pi_\star$.

\subsection{Policy approximation and Deterministic policy gradient} \label{sec:DeterPolGradient}
In most cases, the optimal policy $\vect\pi_\star$ cannot be computed. It is then useful to consider approximations $\vect\pi_{\vect\theta}$ of the optimal policy, carried by a (possibly large) set of parameters $\vect\theta$. The optimal parameters $\vect\theta_\star$ are then given by:
\begin{align}
\vect \theta_\star = \mathrm{arg}\min_{\vect\theta}\, J(\vect\pi_{\vect\theta}) \label{eq:OptParameters}
\end{align}
The gradient associated to the minimization problem \eqref{eq:OptParameters} is referred to as the \textit{deterministic policy gradient} and is given by \cite{Silver2014}:
\begin{align}
\label{eq:DetPiGradient}
\nabla_{\vect{\theta}}\, J(\vect{\pi}_{\vect{\theta}}) = \mathbb{E}_{\vect\pi_{{\vect{\theta}}}}\left[\nabla_{\vect{\theta}} \vect{\pi}_{\vect{\theta}}\, \nabla_{\vect{a}}A_{\vect{\pi}_{\vect{\theta}}}\right],
\end{align}
where $\nabla_{\vect{u}}A_{\vect{\pi}_{\vect{\theta}}}$ is the gradient of the advantage function \eqref{eq:A:definition}. Reinforcement Learning algorithms based on the deterministic policy gradient are forming estimations of \eqref{eq:DetPiGradient} using observed state transitions. The gradient of the advantage function $\nabla_{\vect{a}}A_{\vect{\pi}_{\vect{\theta}}}$ in \eqref{eq:DetPiGradient} is also estimated from the data.

One can observe that for any deterministic policy $\vect\pi_{\vect\theta}$, the advantage function satisfies 
\begin{align}
\label{eq:trivial:A}
A_{\vect{\pi}_{\vect{\theta}}}\left(\vect s, \vect\pi_{\vect\theta}(\vect s)\right) = 0,\qquad \forall \, \vect s,
\end{align}
hence in order to build estimations of the gradient $\nabla_{\vect{u}}A_{\vect{\pi}_{\vect{\theta}}}$, one needs to select inputs $\vect a$ that depart from the deterministic policy $\vect\pi_{\vect\theta}$, so as to be able to observe variations of $A_{\vect{\pi}_{\vect{\theta}}}$, see \eqref{eq:SomeAProperties}, and estimate its gradient. Selecting inputs $\vect a\neq \vect\pi_{\vect\theta}\left(\vect s\right)$ in order to build  the gradient $\nabla_{\vect{u}}A_{\vect{\pi}_{\vect{\theta}}}$ is referred to as \textit{exploration}. 
\subsection{Safe set} \label{sec:SafeSet}
In the following, we will assume the existence of a (possibly) state-dependent \textit{safe set} labelled $\mathbb{S}\left(\vect s\right)\subseteq \mathbb{R}^{n_{\vect a}}$, subset of the input space. See \cite{Zanon2019b,Gros2019a} for similar discussions. The notion of safe set will be used here in the sense that any input selected such that $\vect a\in \mathbb{S}\left(\vect s\right)$ yields safe trajectories with a unitary probability. The construction of the safe set is not the object of this paper. However, we can nonetheless propose some pointer to how such a set is constructed in practice.

Let us consider the constraints $\vect h_\mathrm{s}\left(\vect s,\vect a\right)\leq 0$ describing the subset of the state-input space deemed feasible and safe. Constraints $\vect h$ can include pure state constraints, describing the safe states, pure input constraints, typically describing actuators limitations, and mixed constraints, where the states and inputs are mixed. For the sake of simplicity, we will assume in the following that $\vect h_\mathrm{s}$ is convex.


A common approach to build practical or inner approximations of the safe set $\mathbb{S}\left(\vect s\right)$ is via verifying the safety of an input $\vect a$ explicitly over a finite horizon via predictive control techniques. This verification is based on forming the support of the Markov Process distribution over time, starting from a given state-input pair $\vect s,\vect a$. Consider the set $\vect{X}_+\left(\vect{s},\vect{a}\right)$, support of the state transition \eqref{eq:State:Transition},
\begin{align}
\label{eq:StateTransition:Support}
\vect{X}_+\left(\vect{s},\vect{a}\right) = \left\{\,\left.\vect s_+\,\,\right |\,\, \mathbb{P}\left[\,\vect s_+\,|\vect s,\vect a\right] > 0\,\right\}
\end{align}
Labelling $\vect{X}_k(\vect{s},\vect a,\vect{\pi}^{\mathrm{s}})$ the support of the state of the Markov Process at time $k$, starting from $\vect s,\vect a$ and evolving under policy $\vect{\pi}^{\mathrm{s}}$, the set $\vect{X}_k$ is then given by the recursion:
\begin{align}
\label{eq:SetDispersion}
\vect{X}_k(\vect{s},\vect a,\vect{\pi}^{\mathrm{s}}) &= \vect{X}_+\left(\vect{X}_{k-1},\vect{\pi}^{\mathrm{s}}(\vect{X}_{k-1})\right),
\end{align}
with the boundary condition $ \vect{X}_1 =\vect{X}_+\left(\vect{s},\vect{a}\right)$. An input $\vect a$ is in the safe set $\mathbb{S}\left(\vect s\right)$ if $\vect h_\mathrm{s}\left(\vect s,\vect a\right)\leq 0$ and if there exist a policy $\vect\pi^\mathrm{s}$ such that
\begin{align}
\vect h_\mathrm{s}\left(\vect s_k,\vect\pi_\mathrm{s}\left(\vect s_k\right)\right)\leq 0, \quad \forall\, \vect s_k\in \vect{X}_k(\vect{s},\vect a,\vect{\pi}^{\mathrm{s}}), 
\end{align}
for all $k\geq 1$. This verification is typically performed in practice via scenario trees, tube-based approaches, or direct approximations of the set $\vect{X}_k$ via e.g. ellipsoids or polytopes~\cite{Mayne2014}.

In that context, policy $\vect{\pi}^{\mathrm{s}}$ should ideally be identical to $\vect\pi_{\vect\theta}$. However, for computational reasons, it is typically selected a priori to stabilize the system dynamics, and possibly optimized to minimize the size of the sets $\vect{X}_k$.

Due to the safety requirement, both the policy $\vect\pi_{\vect\theta}$ and the exploration performed by the RL algorithm will have to respect $\vect a\in\mathbb{S}\left(\vect s\right)$, and can therefore not be chosen freely. 

\section{Optimization-based safe policy} \label{sec:NMPCIntro}

In this paper, we will consider parametrized deterministic policies $\vect{\pi}_\theta$ based on parametric Nonlinear Programs (NLPs), and more specifically based on robust NMPC schemes. This approach is formally justified in \cite{Gros2018}. More specifically, we will consider a policy approximation 
\begin{align}
\vect\pi_{\theta} = \vect{u}^\star_0\left(\vect{s},\vect{\theta}\right), \label{eq:PolicyFromNLP}
\end{align}
 where $\vect{u}^\star_0\left(\vect{s},\vect{\theta}\right)$ is the first $n_{\vect a}$ entries of $\vect{u}^\star\left(\vect{s},\vect{\theta}\right)$ generated by the parametric NLP:
\begin{subequations}
\label{eq:Generic:NLP}
\begin{align}
\vect{u}^\star\left(\vect{s},\vect{\theta}\right) = \mathrm{arg}\min_{\vect{u}}&\quad \Phi(\vect{x},\vect{u},\vect{\theta}) \\
\mathrm{s.t.}&\quad \vect f\left(\vect x,\vect u,\vect s,\vect{\theta}\right) = 0, \label{eq:DynamicConstraints}\\
&\quad \vect{h}\left(\vect{x},\vect{u},\vect{\theta}\right) \leq 0. \label{eq:SafetyConstraints}
\end{align}
\end{subequations}
We will then consider that the safety requirement $\vect \pi_{\vect\theta}(\vect s) \in\mathbb{S}(\vect s)$ is imposed via the constraints \eqref{eq:DynamicConstraints}-\eqref{eq:SafetyConstraints}. A special case of \eqref{eq:Generic:NLP} is an optimization scheme in the form:
\begin{subequations}
\label{eq:RobustNMPC:Policy:Constraints:Static}
\begin{align}
\vect u_0^\star\left(\vect s,\vect{\theta}\right) = \mathrm{arg}\min_{\vect u_0}&\quad \Phi(\vect s,\vect u_0,\vect{\theta}) \\
\mathrm{s.t.}&\quad \vect{h}\left(\vect s,\vect u_0,\vect{\theta}\right) \leq 0, \label{eq:SafetyConstraints:Static}
\end{align}
\end{subequations}
where $\vect{h}\leq0$ ought to ensure that $\vect \pi_{\vect\theta}(\vect s) = \vect u_0^\star\left(\vect s,\vect{\theta}\right) \in\mathbb{S}\left(\vect s\right)$. 

While most of the discussions in this paper will take place around the general formulation \eqref{eq:Generic:NLP}, a natural approach to formulate constraints \eqref{eq:DynamicConstraints}-\eqref{eq:SafetyConstraints} such that policy \eqref{eq:PolicyFromNLP} is safe is to build \eqref{eq:Generic:NLP} using robust (N)MPC techniques. 

 \subsection{Policy approximation based on robust NMPC}
The imposition of safety constraints can be treated via robust NMPC approaches. Robust NMPC can take different forms \cite{Mayne2014}, all of which can be eventually cast in the form \eqref{eq:Generic:NLP}. One form of robust robust NMPC schemes is based on scenario trees~\cite{Scokaert1998}, which take the form: 
\begin{subequations}
\label{eq:RobustNMPC:Policy:Constraints:MPC}
\begin{align}
\vect{u}^\star\left(\vect s,\vect{\theta}\right) &=\nonumber\\ \mathrm{arg}\min_{\vect{u}}&\,\, \sum_{j=1}^{N_{\mathrm M}} \left(V_j(\vect x_{j,N},\vect\theta) +  \sum_{k=0}^{N-1} \ell_j(\vect x_{j,k},\vect u_{j,k},\vect\theta)\right)\\
\mathrm{s.t.}&\,\, \vect x_{j,k+1} = \vect F_j\left(\vect x_{j,k},\vect u_{j,k},\vect \theta\right),\,\,\, \vect x_{j,0} = \vect s, \label{eq:Dynamics:MPC:Robust}\\
&\,\,  \vect{h}^\mathrm{s}\left(\vect{x}_{j,k},\vect{u}_{j,k},\vect \theta\right) \leq 0, \label{eq:SafetyConstraints:MPC:Robust} \\
&\,\,  \vect{e}\left(\vect{x}_{j, N},\vect \theta\right) \leq 0, \label{eq:TerminalConstraints:MPC:Robust} \\
&\,\,\, \vect N\left(\vect u\right )= 0, \label{eq:NonAnticipativity}
\end{align}
\end{subequations}
where $\vect F_{1,\ldots,N_{\mathrm M}}$ are the $N_{\mathrm M}$ different models used to support the uncertainty, while $\vect F_0$ is a nominal model supporting the NMPC scheme. Trajectories $\vect x_{j,k}$ and $\vect u_{j,k}$ for $j=1,\ldots,N_\mathrm{M}$ are the different models trajectories and the associated inputs. Functions $\ell_{1,\ldots,N_{\mathrm M}}$, $V_{1,\ldots,N_{\mathrm M}}$ the (possibly different) stage costs and terminal costs applying to the different models. The \textit{non-anticipativity constraints} \eqref{eq:NonAnticipativity} support the scenario-tree structure.
For a given state $\vect s$ and parameters $\vect \theta$, the NMPC scheme \eqref{eq:RobustNMPC:Policy:Constraints:MPC} delivers the input profiles 
\begin{align}
\vect{u}_j^\star\left(\vect s,\vect{\theta}\right) = \left\{\vect u_{j,0}^\star\left(\vect s,\vect{\theta}\right),\ldots, \vect{u}_{j,N}^\star\left(\vect s,\vect{\theta}\right)\right\},\end{align}
with $\vect{u}_{j,i}^\star\left(\vect s,\vect{\theta}\right)\in\mathbb{R}^{n_{\vect a}}$, and \eqref{eq:NonAnticipativity} imposes
\begin{align}
\vect u_0^\star\left(\vect s,\vect{\theta}\right) := \vect u_{i,0}^\star\left(\vect s,\vect{\theta}\right) = \vect u_{j,0}^\star\left(\vect s,\vect{\theta}\right),\quad\forall i,j.
\end{align}
As a result, the NMPC scheme \eqref{eq:RobustNMPC:Policy:Constraints:MPC} generates a parametrized deterministic policy according to:
\begin{align}
\label{eq:Policy0}
\vect{\pi}_\theta\left(\vect s\right) = \vect u_0^\star\left(\vect{s},\vect{\theta}\right) \,\in\,\mathbb{R}^{n_{\vect a}}.
\end{align}
Policy $\vect\pi^\mathrm{s}$ is implicitly deployed in \eqref{eq:RobustNMPC:Policy:Constraints:MPC} via the scenario tree. {If the dispersion set $\vect X_+$ is known, the multiple models $\vect F_{1,\ldots , N_\mathrm{M}}$ and terminal constraints \eqref{eq:TerminalConstraints:MPC:Robust} can be chosen such that the robust NMPC scheme \eqref{eq:RobustNMPC:Policy:Constraints:MPC} delivers $\vect{\pi}_\theta\left(\vect s\right)\in\mathbb{S}\left(\vect s\right)$. Unfortunately, this selection can be difficult in general. We turn next to the robust linear MPC case, where this construction is much simpler.}


\subsection{Safe robust linear MPC} \label{sec:SafeTree}
Exhaustively discussing the construction of the safe scenario tree in \eqref{eq:RobustNMPC:Policy:Constraints:MPC} for a given dispersion set $\vect X_+\left(\vect s,\vect a\right)$ is beyond the scope of this paper. The process can be fairly involved, and we refer to \cite{Scokaert1998,Bernardini2009} for detailed discussions. For the sake of brevity, we will focus on the linear MPC case, whereby the MPC models $\vect F_{1,\ldots,N_{\mathrm M}}$ and policy $\vect\pi^\mathrm{s}$ are linear. 

Let us consider the following outer approximation of  the dispersion set $\vect X_+$:
\begin{align}
\label{eq:DispersionApprox}
\vect X_+\left(\vect s,\vect a\right) \subseteq \vect F_0\left(\vect s,\vect a,\vect \theta\right) + \vect W,\quad \forall \, \vect s,\vect a
\end{align}
where we use a linear nominal model $\vect F_0$ and a polytope $\vect W$ of vertices $\vect W^{1,\ldots,N_{\mathrm M}}$ that can be construed as the extrema of a finite-support process noise, and which can be part (or functions of) the MPC parameters $\vect\theta$. For the sake of simplicity, we assume that $\vect W$ is independent of the state-input pair $\vect s,\vect a$. The models $\vect F_{1,\ldots,N_\mathrm M}$ can then be built based using:
\begin{align}
\vect F_i = \vect F_0 + \vect W^i,\quad i=1\ldots N_{\mathrm M}
\end{align}
and using the linear policy:
\begin{align}
\vect\pi^\mathrm{s}\left(\vect x_{j,k},\vect u_{0,k},\vect x_{0,k}\right) = \vect u_{0,k} - K\left(\vect x_{j,k}-\vect x_{0,k}\right)
\end{align}
where matrix $K$ can be part (or function of) the MPC parameters $\vect\theta$. One can then verify by simple induction that:
\begin{align}
\vect{X}_k(\vect{s},\vect a,\vect{\pi}^{\mathrm{s}}) \subseteq \mathrm{Conv}\left(\vect x_{1,k},\ldots, \vect x_{N_{\mathrm M},k}\right),
\end{align}
for $k=0,\ldots,N+1$, where $ \mathrm{Conv}$ is the convex hull of the set of points $\vect x_{1,k},\ldots, \vect x_{N_{\mathrm M},k}$ solution of the MPC scheme \eqref{eq:RobustNMPC:Policy:Constraints:MPC}. The terminal constraints \eqref{eq:TerminalConstraints:MPC:Robust} ought then be constructed as, e.g., via the Robust Positive Invariant set corresponding to $\vect\pi^\mathrm{s}$ in order to establish safety beyond the MPC horizon. For $\vect{h}^\mathrm{s}$ convex, the MPC scheme \eqref{eq:RobustNMPC:Policy:Constraints:MPC} delivers safe inputs \cite{Mayne2014,Kolmanovsky1998}.

When the dispersion set $\vect X_+\left(\vect s,\vect a\right) $ can only be inferred from data, condition \eqref{eq:DispersionApprox} arguably translates to \cite{Bertsekas1971a}:
\begin{align}
\label{eq:DispersionApprox:Data}
\vect s_{k+1} - \vect F_0\left(\vect s_k,\vect a_k,\vect \theta\right) \in  \vect W,\quad \forall \left(\vect s_{k+1} ,\vect s_k,\vect a_k\right)\in\mathcal D,
\end{align}
where $\mathcal D$ is the set of $N_{\mathcal{D}}$ observed state transitions. Condition \eqref{eq:DispersionApprox:Data} translates into a sample-based condition on the admissible parameters $\vect \theta$, i.e., it speficies the parameters that are safe \textit{with respect to the state transitions observed so far}. Condition \eqref{eq:DispersionApprox:Data} tests whether the points $\vect s_{k+1} - \vect F_0\left(\vect s_k,\vect a_k,\vect \theta\right)$ are in the polytope $\vect W$, which can be easily translated into a set of algebraic constraints imposed on $\vect\theta$. This observation will be used in Section \ref{sec:SafeTree} to build a safe RL-based learning.

We ought to underline here that building $\vect F_0,\,\vect W$ based on \eqref{eq:DispersionApprox:Data} ensures the safety of the robust MPC scheme \eqref{eq:RobustNMPC:Policy:Constraints:MPC} only for an infinitely large, and sufficiently informative data set $\mathcal D$. In practice, using a finite data set entails that safety is ensured with a probability less than 1. The quantification of the probability of having a safe policy for a given, finite data set $\mathcal D$ is beyond the scope of this paper, and is arguably best treated by means of the Information Field Theory \cite{Ensslin2013}. The extension of the construction of a safe MPC presented in this section to the general NMPC case is theoretically feasible, but can be computationally intensive in practice. This aspect of the problem is beyond the scope if this paper.

\section{Safe exploration} \label{sec:SafeExplo}

In this section, we investigate the deployment of the deterministic policy gradient method \cite{Silver2014} when the input space is continuous and restricted by some safety constraints. We will show that in that case the classic tools used in the deterministic policy gradient method need some corrections.

\begin{figure}
	\includegraphics[width=1\linewidth,clip,trim= 75 250 70 370]{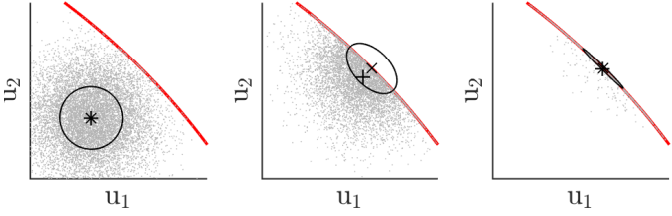} 
	\caption{\footnotesize Illustration of the mean ($+$ symbol) and covariance (ellipsoids) \eqref{eq:MeanAndCov} of the exploration \eqref{eq:ExploDefinition} subject to safety constraints (solid red line). The deterministic policy is depicted as the $\times$ symbol, for the trivial problem \eqref{eq:TrivialExample}, and different values of $\vect \theta_{1,2}$ and $\vect\theta_3=1$. Here the exploration is generated via \eqref{eq:NLP:Disturbed}-\eqref{eq:d:distribution} detailed below. One can observe that $\vect \eta_{\vect e}= 0$ is not possible to achieve when the policy is on the constraints, and that the covariances are impacted by the presence of the constraint. On the right-side graph, one can see how the covariance collapses when $\vect\pi_{\vect\theta}$ strongly activates the constraint.}
	\label{fig:Illustration}
\end{figure}

In order to estimate the gradient of the advantage function $\nabla_{\vect{a}}A_{\vect{\pi}_{\vect{\theta}}}$, the inputs $\vect{a}$ applied to the real system must differ from the actual policy $\vect{\pi}_{\vect{\theta}}\left(\vect s\right)$, such that the advantage function $\hat A^{\vect{w}}_{\vect{\pi}_{\vect{\theta}}}\left(\vect s,\vect{a}\right) $ is not trivially zero on the system trajectories, see \eqref{eq:trivial:A}. The exploration 
\begin{align}
\label{eq:ExploDefinition}
\vect{e} := \vect{a} - \vect{\pi}_{\vect{\theta}}(\vect s),
\end{align}
is typically generated via selecting the inputs $\vect a$ using a stochastic policy $\pi_{\vect\theta}^\sigma[\,\vect a\,|\,\vect s\,]$, where $\sigma$ relates to its covariance, having most of its mass in a neighborhood of $\vect{\pi}_{\vect{\theta}}(\vect s)$. In the following, we will need the conditional mean and covariance of $\vect e$:
\begin{subequations}
\label{eq:MeanAndCov}
\begin{align}
\vect{\eta}_{\vect{e}}\left(\vect s\right) &= \mathbb{E}\left[\vect{e}\,|\,\vect s \right],\\ \Sigma_{\vect{e}}\left(\vect s\right) &= \mathbb{E}\left[\left.\left(\vect e - \vect{\eta}_{\vect{e}}\right)\left(\vect e - \vect{\eta}_{\vect{e}}\right)^\top\,\right|\,\vect s \right].
\end{align}
\end{subequations}
The restriction of the exploration $\vect e$ to yield inputs $\vect a$ in the safe set $\mathbb{S}\left(\vect s\right)$ can cause the exploration $\vect e$ to not be centred, i.e., $\vect\eta_{\vect e}= 0$ may not hold, and the covariance $\Sigma_{\vect{e}}\left(\vect s\right)$ can be restricted by the safe set. This observation is illustrated in Fig. \ref{fig:Illustration}, where the trivial static problem:
\begin{subequations}
\label{eq:TrivialExample}
\begin{align}
\vect \pi_{\vect\theta} = \mathrm{arg}\min_{\vect u}&\quad \frac{1}{2}\left\|\vect u - \matr{c}{\vect\theta_1\\\vect\theta_2}\right\|^2\\
\mathrm{s.t.}&\quad \left\|\vect u\right\|^2 \leq \vect\theta_3 \label{eq:TrivialExample:Const}
\end{align}
\end{subequations}
was used, and the exploration generated via \eqref{eq:NLP:Disturbed}-\eqref{eq:d:distribution} detailed below. 

The fact $\vect\eta_{\vect e}$ and $\Sigma_{\vect e}$ cannot be fully chosen in the presence of constraints ought to be accounted for when forming estimations of the gradient of the advantage function $\nabla_{\vect{a}}A_{\vect{\pi}_{\vect{\theta}}}$ in order to avoid biasing the estimation of the policy gradient \eqref{eq:DetPiGradient}. We develop next conditions on the exploration such that $\nabla_{\vect{a}}A_{\vect{\pi}_{\vect{\theta}}}$ can be estimated correctly.

\subsection{Estimation of $\nabla_{\vect{a}}A_{\vect{\pi}_{\vect{\theta}}}$}

A difficulty arises here when forming estimations of $\nabla_{\vect{a}}A_{\vect{\pi}_{\vect{\theta}}}$ using restricted explorations, which we detail hereafter. It is well known that estimating $\nabla_{\vect{a}}A_{\vect{\pi}_{\vect{\theta}}}$ directly is very difficult, hence one typically considers estimating the advantage function $A_{\vect{\pi}_{\vect{\theta}}}$ instead, from which the gradient $\nabla_{\vect{a}}A_{\vect{\pi}_{\vect{\theta}}}$ is evaluated. The estimation of the advantage function is carried by the function approximator $\hat A^{\vect{w}}_{\vect{\pi}_{\vect{\theta}}}$, parametrized by $\vect w$. One then seeks a solution to the least-squares problem \cite{Silver2014}:
\begin{align}
\label{eq:A:TDEstimation}
\vect{w} =\mathrm{arg}\min_{\vect{w}}\, \frac{1}{2}\mathbb{E}_{\pi^\sigma_{{\vect{\theta}}}}\left[\left(Q_{\vect{\pi}_{\vect{\theta}}} - \hat V_{\vect{\pi}_{\vect{\theta}}} - \hat A^{\vect{w}}_{\vect{\pi}_{\vect{\theta}}}   \right)^2\right],
\end{align}
where the value function estimation $\hat V_{\vect{\pi}_{\vect{\theta}}}\approx V_{\vect{\pi}_{\vect{\theta}}}$ is a baseline supporting the evaluation of $\vect{w}$. Temporal-Difference or Monte Carlo techniques \cite{Sutton1999} are typically used to tackle \eqref{eq:A:TDEstimation}. The policy gradient is then evaluated as:
\begin{align}
\label{eq:DetPiGradient:Approx}
\widehat{\nabla_{\vect{\theta}}\, J(\vect{\pi}_{\vect{\theta}})} &= \mathbb{E}_{\pi^\sigma_{{\vect{\theta}}}}\left[\nabla_{\vect{\theta}} \vect{\pi}_{\vect{\theta}}\, \nabla_{\vect{a}}\hat A^{\vect{w}}_{\vect{\pi}_{\vect{\theta}}}\right].
\end{align}

A compatible linear approximator $\hat A^{\vect{w}}_{\vect{\pi}_{\vect{\theta}}}$ is typically preferred, in order for \eqref{eq:DetPiGradient:Approx} to match \eqref{eq:DetPiGradient}. In this paper, we propose to use the following function approximator inspired from \cite{Silver2014}: 
\begin{align}
\label{eq:Aparam:Compatible}
\hat A^{\vect{w}}_{\vect{\pi}_{\vect{\theta}}}\left(\vect s,\vect{a}\right) =  \vect w^\top \nabla_{\vect{\theta}}\vect{\pi}_{\vect{\theta}} M\left(\vect{a} - \vect{\pi}_{\vect{\theta}}-\vect c\right),
\end{align}
where $M$ and $\vect c$ are a (possibly) state-dependent $\mathbb{R}^{{n_{\vect a}}\times {n_{\vect a}}}$ symmetric matrix and $\mathbb{R}^{n_{\vect a}}$ vector. In \cite{Silver2014}, $M=I$ and $\vect c=0$ is used but we will show in the following that we need in principle to make a different choice when the input is restricted to $\mathbb{S}(\vect s)$. The following proposition delivers general conditions in order for \eqref{eq:DetPiGradient:Approx} to match \eqref{eq:DetPiGradient} when the exploration is restricted. Before delivering the Proposition, let us establish some key assumptions.
\vspace{0.25cm}
\begin{Assumption}\label{ass:Assumptions} The following hold:
\begin{enumerate}[label=\alph*.]
\item \label{ass:Assumptions:1} $Q_{\vect{\pi}_{\vect{\theta}}}\left(\vect s,\vect a\right)$ is almost everywhere at least twice differentiable in $\vect{a}$ on $\mathbb{S}(\vect{s})$ for all feasible $\vect{s}$, and its gradient with respect to $\vect{a}$ are polynomially bounded.
\item Stability assumption \eqref{ass:Stability} holds.
\item \label{ass:Assumptions:4} The MDP state probability density is bounded, i.e. it can not hold Dirac-like densities.
\item \label{ass:Assumptions:5} $\vect e$ results from the transformation of a Normal distribution via a polynomially bounded function
\item \label{ass:Assumptions:6} The following limits:
\begin{subequations}
\label{eq:UnbiasedPiGradientConditions}
\begin{align}
&\lim_{\sigma\rightarrow 0}\frac{1}{\sigma} \left(\vect\eta_{\vect e}-\vect c \right) = 0
\label{eq:UnbiasedPiGradientCondition:Mean} \\
&\lim_{\sigma\rightarrow 0} \frac{1}{\sigma}M\Sigma_{\vect{e}}=I \label{eq:UnbiasedPiGradientCondition:Var}
\end{align}
\end{subequations}
hold for all feasible $\vect s$.
\end{enumerate}
\end{Assumption}
\textit{Remark:}  We ought to underline here that assumptions \ref{ass:Assumptions}\ref{ass:Assumptions:1}-\ref{ass:Assumptions}\ref{ass:Assumptions:5} are typically needed in RL algorithms, though often not explicitly stated. Assumption \ref{ass:Assumptions}\ref{ass:Assumptions:6} is not standard, and relates specifically to the problem of having restricted exploration. It essentially requires $\vect c$ to be an asymptotic estimation of the exploration mean $\vect\eta_{\vect e}$ and $M$ to be a (scaled) asymptotic estimaton of the inverse of the exploration covariance $\Sigma_{\vect e}$. We should additionally observe that when the exploration is centred and isotropically distributed (i.e. $\vect\eta_{\vect e}=0,\,\Sigma_{\vect e} = \sigma I$), then $M=I,\, \vect c=0$ satisfy \eqref{eq:UnbiasedPiGradientConditions}, and the results found in \cite{Silver2014} hold.

\begin{Proposition} \label{Prop:ConditionForDetPolicy} Under Assumption \ref{ass:Assumptions}, the deterministic policy gradient estimation \eqref{eq:DetPiGradient:Approx} is asymptotically exact, i.e.
\begin{align}
\label{eq:Exact:PolicyGradient}
\lim_{\sigma\rightarrow 0 }\widehat{\nabla_{\vect{\theta}}\, J(\vect{\pi}_{\vect{\theta}})} =  {\nabla_{\vect{\theta}}\, J(\vect{\pi}_{\vect{\theta}})} .
\end{align}
\end{Proposition}

\begin{IEEEproof} Using \eqref{eq:Aparam:Compatible}, the solution of the least-squares problem \eqref{eq:A:TDEstimation} satisfies the stationarity condition: 
\begin{align}
\label{eq:Q:Fitting:Stat}
\mathbb{E}_{\pi^\sigma_{{\vect{\theta}}}}\left[\nabla_{\vect{\theta}}\vect{\pi}_{\vect{\theta}}M\left(\vect{e}-\vect c\right)\left(Q_{\vect{\pi}_{\vect{\theta}}} - \hat V_{\vect{\pi}_{\vect{\theta}}} - \hat A^{\vect{w}}_{\vect{\pi}_{\vect{\theta}}}  \right)\right] = 0 .
\end{align}
Since $Q_{\vect{\pi}_{\vect{\theta}}}$ is at least twice differentiable almost everywhere, its second-order expansion in $\vect{a}$ at $\vect{e}=0$ is valid almost everywhere, i.e.
\begin{align}
Q_{\vect{\pi}_{\vect{\theta}}}\left(\vect s,\vect a\right) &= \left(V_{\vect{\pi}_{\vect{\theta}}}+\nabla_{\vect a} Q_{\vect{\pi}_{\vect{\theta}}}^\top\left(\vect a - \vect\pi_{\vect \theta}\right) + \vect\xi\right)_{\vect s,\vect a = \vect\pi_{\vect \theta}} \nonumber \\
&= V_{\vect{\pi}_{\vect{\theta}}}\left(\vect s\right) + \nabla_{\vect a} A_{\vect{\pi}_{\vect{\theta}}}\left(\vect s,\vect\pi_{\vect \theta}\right)^\top\vect e + \vect\xi,
\end{align}
where $\vect\xi$ is the second-order remainder of the Taylor expansion of $Q_{\vect{\pi}_{\vect{\theta}}}$, and where we use $\nabla_{\vect a} Q_{\vect{\pi}_{\vect{\theta}}} = \nabla_{\vect a} A_{\vect{\pi}_{\vect{\theta}}}$, see \eqref{eq:A:definition}. Because $Q_{\vect{\pi}_{\vect{\theta}}}$ is twice differentiable almost everywhere and using \ref{ass:Assumptions}\ref{ass:Assumptions:4}, $\vect\xi$ is of order $\mathcal O(\|\vect e\|^2)$ almost everywhere. We then observe that \eqref{eq:Q:Fitting:Stat} becomes:
\begin{align}
\label{eq:Q:Fitting:Bias}
&\mathbb{E}_{\pi^\sigma_{{\vect{\theta}}}}\left[\nabla_{\vect{\theta}}\vect{\pi}_{\vect{\theta}}M\left(\vect{e}-\vect c\right)\vect{e}^\top\left(\nabla_{\vect{a}}A_{\vect{\pi}_{\vect{\theta}}}-\nabla_{\vect{a}}\hat A^{\vect{w}}_{\vect{\pi}_{\vect{\theta}}}\right) \right]\\&\hspace{0.5cm}+\mathbb{E}_{\pi^\sigma_{{\vect{\theta}}}}\left[ \nabla_{\vect{\theta}}\vect{\pi}_{\vect{\theta}}M\left(\vect{e}-\vect c\right)\vect\xi\right]\nonumber\\&\hspace{1cm}+\mathbb{E}_{\pi^\sigma_{{\vect{\theta}}}}\left[ \nabla_{\vect{\theta}}\vect{\pi}_{\vect{\theta}}M\left(\vect{e}-\vect c\right)\left(V_{\vect{\pi}_{\vect{\theta}}} -\hat V_{\vect{\pi}_{\vect{\theta}}}\right) \right] = 0.\nonumber
\end{align}
Function $\vect \xi$ is of second-order or more in $\vect e$ and using the second assumption, it is polynomially bounded. Moreover, using assumption \ref{ass:Assumptions}\ref{ass:Assumptions:5}, and using arguments from the Delta method \cite{Oehlert1992}, we can conclude that: 
\begin{align}
\lim_{\sigma\rightarrow 0}\frac{1}{\sigma}\mathbb{E}\left[ \, \nabla_{\vect{\theta}}\vect{\pi}_{\vect{\theta}}M \left(\vect{e}-\vect c\right)\vect\xi\,|\, \vect s\right] =0,\quad \forall\,\vect s \label{eq:ThrirdOrderVanishes}
\end{align}
holds. It follows that the second term of \eqref{eq:Q:Fitting:Bias} asymptotically vanishes faster than $\sigma$. Moreover, \eqref{eq:UnbiasedPiGradientCondition:Mean} guarantees that the third term of \eqref{eq:Q:Fitting:Bias} also asymptotically vanishes faster than $\sigma$. It follows that
\begin{align}
\label{eq:Q:Fitting:Bias:Repeat}
\lim_{\sigma\rightarrow 0}\frac{1}{\sigma}\mathbb{E}_{\pi^\sigma_{{\vect{\theta}}}}\left[\nabla_{\vect{\theta}}\vect{\pi}_{\vect{\theta}}M\left(\vect{e}-\vect c\right)\vect{e}^\top\left(\nabla_{\vect{a}}A_{\vect{\pi}_{\vect{\theta}}}-\nabla_{\vect{a}}\hat A^{\vect{w}}_{\vect{\pi}_{\vect{\theta}}}\right) \right] = 0.
\end{align}
Using \eqref{eq:UnbiasedPiGradientConditions} we observe that using \eqref{eq:UnbiasedPiGradientCondition:Mean}:
\begin{align}
&\lim_{\sigma\rightarrow 0}\frac{1}{\sigma}\mathbb{E}\left[M\left(\vect{e}-\vect c\right)\vect{e}^\top \right]\label{eq:Bla} \\&= \lim_{\sigma\rightarrow 0}\frac{1}{\sigma}M\left(\Sigma_{\vect{e}} + \vect\eta_{\vect e} \vect\eta_{\vect e}^\top -\vect c \vect\eta_{\vect e}^\top\right) =  \lim_{\sigma\rightarrow 0}\frac{1}{\sigma}M\Sigma_{\vect{e}}\nonumber.
\end{align}
We finally conclude that \eqref{eq:Q:Fitting:Stat} and \eqref{eq:Q:Fitting:Bias:Repeat} with \eqref{eq:DetPiGradient:Approx} entail that
\begin{align}
&\lim_{\sigma\rightarrow 0}\frac{1}{\sigma}\mathbb{E}_{\pi^\sigma_{\vect{\theta}}}\left[\nabla_{\vect{\theta}}\vect{\pi}_{\vect{\theta}}M\Sigma_{\vect{e}}\left(\nabla_{\vect{a}}A_{\vect{\pi}_{\vect{\theta}}}-\nabla_{\vect{a}}\hat A^{\vect{w}}_{\vect{\pi}_{\vect{\theta}}}\right) \right]\label{eq:CovarianceConditionForExactGradient}\\
&= \mathbb{E}_{{\vect{\pi}_{\vect{\theta}}}}\left[\nabla_{\vect{\theta}}\vect{\pi}_{\vect{\theta}}\left(\nabla_{\vect{a}}A_{\vect{\pi}_{\vect{\theta}}}-\nabla_{\vect{a}}\hat A^{\vect{w}}_{\vect{\pi}_{\vect{\theta}}}\right) \right] = 0.\nonumber
\end{align}
where assumption \eqref{ass:Stability} yields asymptotically the equivalence between $\mathbb{E}_{\pi^\sigma_{\vect{\theta}}}\left[.\right]$ and $\mathbb{E}_{{\vect{\pi}_{\vect{\theta}}}}\left[.\right]$. Equation \eqref{eq:Exact:PolicyGradient} follows. 
\end{IEEEproof}
\vspace{0.25cm}
We now turn to proposing a computationally effective method to generate a safe exploration and to compute mean and covariance estimations for $\vect e$, i.e., a matrix $M(\vect s)$ and vector $\vect c(\vect s)$ that satisfy conditions \eqref{eq:UnbiasedPiGradientConditions}.
\section{Optimization-based safe exploration} \label{eq:Numerics}

In this section we proposed a modification of \eqref{eq:Generic:NLP} allowing one to build a stochastic policy $\pi_{\vect\theta}^\sigma$ that produces safe inputs for exploration, and for which the corrections $M$ and $\vect c$ satisfying Assumption~\ref{ass:Assumptions}\ref{ass:Assumptions:6} are cheap to compute. The proposed approach will use the primal-dual interior-point method and techniques from parametric Nonlinear Programming. To that end we will consider inputs $\vect a = \vect{u}_0^\mathrm{d}\left(\vect s,\vect\theta,\vect d\right)$ generated from:
\begin{subequations}
\label{eq:NLP:Disturbed}
\begin{align}
 \vect{u}^\mathrm{d}\left(\vect s,\vect\theta,\vect d\right) = \mathrm{arg}\min_{\vect{u}}&\quad\Phi^{\vect d}(\vect{x},\vect u,\vect{\theta},\vect d)\\
\mathrm{s.t.}&\quad \vect f\left(\vect x,\vect u,\vect s,\vect{\theta}\right) = 0, \\
&\quad \vect{h}\left(\vect{x},\vect{u},\vect{\theta}\right) \leq 0, \label{eq:NLP:Disturbed:Ineq}
\end{align}
\end{subequations}
where $\Phi^{\vect d}(\vect{u},\vect s,\vect{\theta},\vect d)$ is a modified version of the cost function $\Phi$ in \eqref{eq:Generic:NLP}, 
and $\vect d\in\mathbb{R}^{n_{\vect a}}$ is drawn from a Normal, centred probability distribution of density:
\begin{align}
\vect d\sim \mathcal N\left(0,\sigma \Sigma\left(\vect s\right)\right) \label{eq:d:distribution}
\end{align}
of covariance $\sigma \Sigma(\vect s)$, where $\Sigma$ (possibly) depends on $\vect s$. The stochastic policy $\pi_{\vect\theta}^\sigma$ will then result from \eqref{eq:NLP:Disturbed}-\eqref{eq:d:distribution}. A simple choice for $\Phi^{\vect d}(\vect{x},\vect u,\vect{\theta},\vect d)$ is a gradient disturbance:
\begin{align}
\label{eq:CostGradientDisturbance}
\Phi^{\vect d}(\vect{u},\vect s,\vect{\theta},\vect d) = \Phi(\vect{x},\vect u,\vect{\theta},\vect d) + \vect d^\top\vect u_0.
\end{align}
One can verify that $\vect a = \vect{u}_0^\mathrm{d}\left(\vect s,\vect\theta,\vect d\right) \in\mathbb{S}\left(\vect s\right)$ by construction, such that the exploration is safe. 

Deploying the principles detailed in Section \ref{sec:DeterPolGradient} and Proposition \ref{Prop:ConditionForDetPolicy} requires one to form at each time step asymptotically accurate estimations $\vect c$, $M$ of the mean $\vect\eta_{\vect e}$ and covariance $\Sigma_{\vect e}$. For $\vect e$ restricted to generate inputs in a non-trivial safe set $\mathbb{S}(\vect s)$, estimating $\vect\eta_{\vect e}$ and $\Sigma_{\vect e}$ requires in general sampling the distribution of $\vect e$ generated by \eqref{eq:NLP:Disturbed}-\eqref{eq:d:distribution}, which is unfortunately computationally expensive as a large number of sample is required and each sample requires solving \eqref{eq:NLP:Disturbed}. 

An alternative to estimating $\vect\eta_{\vect e}$ and $\Sigma_{\vect e}$ using sampling is to form these estimations via a Taylor expansion of $\vect{u}_0^\mathrm{d}\left(\vect s,\vect\theta,\vect d\right)$ in $\vect d$. Unfortunately, $\vect{u}_0^\mathrm{d}\left(\vect s,\vect\theta,\vect d\right)$ is in general non-smooth due to the presence of inequality constraints in \eqref{eq:NLP:Disturbed}. To alleviate this problem, in this section, we propose to cast \eqref{eq:NLP:Disturbed} in a primal-dual interior point formulation, i.e., we consider that the solutions of \eqref{eq:NLP:Disturbed} are obtained from solving the relaxed Karush-Kuhn-Tucker (KKT) conditions \cite{Biegler2010}:
\begin{align}
\label{eq:NLP:Disturbed:IP}
\vect r_\tau(\vect z,\vect \theta,\vect d) = \matr{c}{\nabla_{\vect w}\Phi^\mathrm{d} + \nabla_{\vect w}\vect h \vect \mu + \nabla_{\vect w}\vect f \vect \lambda  \\
\vect f \\
\mathrm{diag}(\vect\mu) \vect h+\tau} = 0,
\end{align}
for $\tau > 0$, and under the conditions $\vect h < 0$ and $\vect\mu > 0$. Here $\vect \mu,\vect \lambda$ are the multipliers associated to the equality and inequality constraints in \eqref{eq:NLP:Disturbed}, and we label $\vect w = \left\{\vect x,\vect u\right\}$ and $\vect z = \left\{\vect w,\,\vect\lambda,\vect\mu\right\}$ the primal-dual variables of \eqref{eq:NLP:Disturbed:IP}. We will label $\vect{u}^{\tau}\left(\vect s,\vect{\theta},\vect{d}\right)$ the parametric primal solution of \eqref{eq:NLP:Disturbed:IP}. 

The error between the true solution of \eqref{eq:NLP:Disturbed} and the one delivered by solving \eqref{eq:NLP:Disturbed:IP} is of the order of the relaxation parameter $\tau$, and the solution $\vect{u}^{\tau}\left(\vect s,\vect{\theta},\vect{d}\right)$ is guaranteed to satisfy the constraints of \eqref{eq:NLP:Disturbed} for all $\tau\geq 0$, hence \eqref{eq:NLP:Disturbed:IP} delivers safe policies if  \eqref{eq:NLP:Disturbed} does. The relaxed KKT conditions \eqref{eq:NLP:Disturbed:IP} yield a smooth function $\vect{u}^{\tau}\left(\vect s,\vect{\theta},\vect{d}\right)$, such that its Taylor expansion is well-defined everywhere. The relaxed KKT conditions \eqref{eq:NLP:Disturbed:IP} will be used next to generate the safe exploration, and the deterministic policy: 
\begin{align}
\vect \pi_{\vect\theta}^\tau = \vect{u}_0^{\tau}\left(\vect s,\vect{\theta},\vect{0}\right). \label{eq:pitau:def}
\end{align}

\subsection{Covariance and mean estimators}


For the sake of clarity, let us use the short notation $\vect g\left(\vect s,\vect\theta,\vect d\right) := \vect{u}_0^{\tau}\left(\vect s,\vect{\theta},\vect{d}\right)$. Hence $\vect g$ is evaluated by solving \eqref{eq:NLP:Disturbed:IP} and extracting the first control input $\vect u^\tau_0\in\mathbb{R}^{n_{\vect a}}$. This function will be instrumental in building cheap mean and covariance estimators satisfying \eqref{eq:UnbiasedPiGradientConditions}. Let us provide these estimators in the following Proposition.

\begin{Proposition} If $ \vect{u}^\mathrm{d}\left(\vect s,\vect\theta,\vect d\right) $ arising from NLP \eqref{eq:NLP:Disturbed} is polynomially bounded in $\vect d$, then the following mean and covariance estimators:
\begin{subequations}
\label{eq:CovMean:Estimators}
\begin{align}
\vect c &= \frac{1}{2}\sum_{i,j=1}^{n_{\vect a}}\frac{\partial^2 \vect g}{\partial \vect d_i\vect d_j} \Sigma_{ij}, 
 \label{eq:MeanEstimator}\\
M&=  \left(\frac{\partial\vect g}{\partial \vect d}\Sigma\frac{\partial\vect g}{\partial \vect d}^\top\right)_{\vect d=0}^{-1}  \label{eq:CovEstimator}
\end{align}
\end{subequations}
satisfy conditions \eqref{eq:UnbiasedPiGradientConditions}, where $\Sigma$ is used in \eqref{eq:d:distribution}. 
\end{Proposition}
\begin{IEEEproof} We observe that:
\begin{align}
\label{eq:MeanEstimation0}
\vect\eta_{\vect e} &= \mathbb{E}\left[\vect g\left(\vect s,\vect\theta,\vect d\right) - \vect g\left(\vect s,\vect\theta,0\right)\right] \\
&=\mathbb{E}\left[\left.\frac{\partial\vect g}{\partial \vect d}\right|_{\vect d=0}\vect d + \frac{1}{2}\sum_{i,j=1}^{n_{\vect a}}\frac{\partial^2 \vect g}{\partial \vect d_i\vect d_j} \vect d_i\vect d_j + \vect \varsigma\right] \nonumber,
\end{align}
where $\vect\varsigma$ is the third-order remainder of the expansion of $\vect g$. We observe that since $\vect d$ has zero mean \eqref{eq:MeanEstimation0} becomes:
\begin{align}
\label{eq:MeanEstimation1}
\vect\eta_{\vect e} &= \frac{\sigma}{2}\sum_{i,j=1}^{n_{\vect a}}\frac{\partial^2 \vect g}{\partial \vect d_i\vect d_j} \Sigma_{ij} +\mathbb{E}\left[ \vect \varsigma\right] .
\end{align}
We also observe that $\mathbb{E}\left[ \vect \varsigma\right] = \mathcal O(\sigma^2)$ holds using arguments from the Delta method \cite{Oehlert1992}. It follows that \eqref{eq:MeanEstimator} satisfies \eqref{eq:UnbiasedPiGradientCondition:Mean}. Furthermore, we observe that
\begin{align}
\Sigma_{\vect e} = \mathbb{E}\left[\vect e \vect e^\top\right] - \vect\eta_{\vect e}\vect\eta_{\vect e}^\top,
\end{align}
and that:
\begin{align}
\label{eq:CovEstimation}
 \mathbb{E}\left[\vect e \vect e^\top\right] = \sigma \left.\frac{\partial\vect g}{\partial \vect d}\Sigma\frac{\partial\vect g}{\partial \vect d}^\top\right|_{\vect d=0} + \mathcal O(\sigma^2),
\end{align}
holds using similar arguments as for \eqref{eq:MeanEstimation0}-\eqref{eq:MeanEstimation1}. It follows that:
\begin{align}
\lim_{\sigma\rightarrow 0}\frac{1}{\sigma}M\Sigma_{\vect{e}} = \left(\frac{\partial\vect g}{\partial \vect d} \Sigma\frac{\partial\vect g}{\partial \vect d}^\top\right)^{-1}\left(  \frac{\partial\vect g}{\partial \vect d} \Sigma\frac{\partial\vect g}{\partial \vect d}^\top \right) = I, \label{eq:MSigmae}
\end{align}
where the Jacobians are evaluated at $\vect d=0$. 
\end{IEEEproof}
\vspace{0.25cm}
Note that deploying \eqref{eq:CovEstimator} requires the Jacobian $\frac{\partial\vect g}{\partial \vect d}\in\mathbb{R}^{n_{\vect a}\times n_{\vect a}}$ to be full rank. This Jacobian also appears in \cite{Gros2019a} to develop the stochastic policy gradient counterpart of this paper, and its rank is investigated. We will not repeat in detail this analysis here, but let us recall its conclusion: for the choice of cost function \eqref{eq:CostGradientDisturbance}, the Jacobian $\frac{\partial\vect g}{\partial \vect d}$ is full rank for any $\tau >0$ if the NLP \eqref{eq:NLP:Disturbed} satisfies the Linearly Independent Constraint Qualification (LICQ) and the Second-Order Sufficient Condition (SOSC). However, $\frac{\partial\vect g}{\partial \vect d}$ can tend to a rank deficient matrix for $\tau \rightarrow 0$ if $ \vect{u}^\mathrm{d}_0$ delivered by \eqref{eq:NLP:Disturbed} activates some of the inequality constraints \eqref{eq:NLP:Disturbed:Ineq}.

\begin{figure}
	\includegraphics[width=1\linewidth,clip,trim= 70 225 70 310]{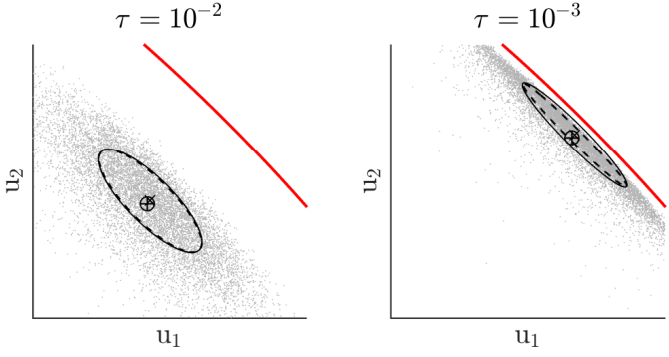} 
	\caption{\footnotesize Illustration of the mean and covariance \eqref{eq:MeanAndCov} of the exploration \eqref{eq:ExploDefinition} subject to safety constraints (solid red line here), generated by the interior-point approach \eqref{eq:NLP:Disturbed:IP} for two value of the relaxation parameter $\tau$. The policy is generated by \eqref{eq:TrivialExample}, and the exploration by \eqref{eq:NLP:Disturbed}-\eqref{eq:CostGradientDisturbance}, with $\Sigma = I$. The mean estimator $\vect c$ (+ symbol) and covariance estimator $M^{-1}$ (dashed ellipsoid) \eqref{eq:CovMean:Estimators} are compared to the ones estimated by sampling (o symbol, solid line ellipsoid).}
	\label{fig:Illustration2}
\end{figure}

Similarly to what has been reported in \cite{Gros2019a}, this issue disappears in some specific cases, which are discussed in the following Proposition, and illustrated in Fig. \ref{fig:Illustration3} below.

\begin{Proposition}\label{eq:DetPolicy:WellDefined} For the choice of cost function \eqref{eq:CostGradientDisturbance}, and if the MPC model dynamics and constraints are not depending on the parameters, i.e. $\nabla_{\vect\theta}\vect h=0,\,\nabla_{\vect\theta}\vect f=0 $, then the choice of $M,\,\vect c$ proposed by \eqref{eq:CovMean:Estimators}  yields \eqref{eq:Exact:PolicyGradient} for $\tau\rightarrow 0$ if problem \eqref{eq:NLP:Disturbed} fulfils LICQ and SOSC.
\end{Proposition}

\begin{IEEEproof}We will prove this statement in an active-set setting deployed on \eqref{eq:NLP:Disturbed}, with \eqref{eq:CovEstimator} evaluated via a pseudo-inverse. The statement of the Proposition will then hold from the convergence of the Interior-Point solution to the active-set one. We will then investigate \eqref{eq:Bla}-\eqref{eq:CovarianceConditionForExactGradient} in that context. Using \eqref{eq:CovEstimator} with a pseudo-inverse, and using similar developments as in Proposition \eqref{Prop:ConditionForDetPolicy}, one can verify that:
\begin{align}
&\lim_{\sigma\rightarrow 0}\frac{1}{\sigma}\mathbb{E}\left[\nabla_{\vect{\theta}}\vect{\pi}_{\vect{\theta}}M\left(\vect{e}-\vect c\right)\vect{e}^\top\right] \\&\qquad =\nabla_{\vect{\theta}}\vect{\pi}_{\vect{\theta}}\left(\frac{\partial\vect g}{\partial \vect d}\Sigma\frac{\partial\vect g}{\partial \vect d}^\top\right)^{+}\left(\frac{\partial\vect g}{\partial \vect d}\Sigma\frac{\partial\vect g}{\partial \vect d}^\top\right)\nonumber \\
&\qquad= \left(\frac{\partial\vect g}{\partial \vect d}\Sigma\frac{\partial\vect g}{\partial \vect d}^\top\right)^{+}\left(\frac{\partial\vect g}{\partial \vect d}\Sigma\frac{\partial\vect g}{\partial \vect d}^\top\right) \frac{\partial\vect g}{\partial \vect \theta}. \nonumber
\end{align}
We will then prove that under the assumptions of this Proposition, $\frac{\partial\vect g}{\partial \vect \theta}$ is in the range space of $\frac{\partial\vect g}{\partial \vect d}\Sigma\frac{\partial\vect g}{\partial \vect d}^\top$, such that
\begin{align}
\label{eq:PinvCompatible}
\left(\frac{\partial\vect g}{\partial \vect d}\Sigma\frac{\partial\vect g}{\partial \vect d}^\top\right)^{+}\left(\frac{\partial\vect g}{\partial \vect d}\Sigma\frac{\partial\vect g}{\partial \vect d}^\top\right) \frac{\partial\vect g}{\partial \vect \theta}  =  \frac{\partial\vect g}{\partial \vect \theta} 
\end{align}
holds. To that end, consider $\mathbb{A}$ the (strictly) active set of \eqref{eq:NLP:Disturbed}, i.e., the set of indices $i$ such that $\vect h_i=0,\vect\mu_i >0$ at the solution. We observe that 
\begin{align}
\matr{cc}{H& \nabla_{\vect w}\vect q \\
\nabla_{\vect w}\vect q^\top & 0}\matr{c}{\frac{\partial \vect w}{\partial \vect d} \\ \frac{\partial \vect \nu}{\partial \vect d}} =- \matr{c}{\nabla_{\vect w\vect d}\Phi^\mathrm{d} \\ 0},
\end{align}
where $H$ is the Hessian of the Lagrange function associated to \eqref{eq:NLP:Disturbed} and
\begin{align}
\vect{q} =  \matr{c}{\vect f \\ \vect h_{\mathbb{A}}},\qquad \vect\nu=\matr{c}{\vect\lambda\\ \vect\mu_{\mathbb{A}}}.
\end{align}
Defining $\mathcal N_{\mathbb{A}}$ the null space of $\nabla_{\vect w}\vect q^\top $, i.e. $\nabla_{\vect w}\vect q^\top\mathcal N_{\mathbb{A}}=0$, we observe that:
\begin{align}
\frac{\partial \vect g}{\partial \vect d}  =- \mathcal N_{\mathbb{A}_0}\left(\mathcal N_{\mathbb{A}}^\top H \mathcal N_{\mathbb{A}}\right)^{-1}\mathcal N_{\mathbb{A}_0}^\top,
\end{align}
where $\mathcal{N}_{\mathbb{A}_0} = \matr{cccc}{I_{{n_{\vect a}}\times {n_{\vect a}}} & 0 & \ldots & 0}\mathcal{N}_{\mathbb{A}}$.
Using a similar reasoning, and since $\frac{\partial \nabla_{\vect w}\vect q}{\partial \vect \theta}=0$, we observe that: 
\begin{align}
\frac{\partial \vect g}{\partial \vect \theta} = - \mathcal N_{\mathbb{A}_0}\left(\mathcal N_{\mathbb{A}}^\top H \mathcal N_{\mathbb{A}}\right)^{-1}\mathcal N_{\mathbb{A}}^\top \nabla_{\vect w\vect\theta}\Phi,
\end{align}
such that $\frac{\partial \vect g}{\partial \vect \theta}$ is in the range space of $\frac{\partial \vect g}{\partial \vect d}$. As a result, for $\Sigma$ full rank $\frac{\partial \vect g}{\partial \vect \theta}$ is in the range space of $\frac{\partial\vect g}{\partial \vect d}\Sigma\frac{\partial\vect g}{\partial \vect d}^\top$, such that \eqref{eq:PinvCompatible} holds. Using \eqref{eq:CovEstimator} defined via the pseudo-inverse, and \eqref{eq:Bla}, \eqref{eq:MeanEstimation1}-\eqref{eq:CovEstimation} one can observe that:
\begin{align}
\label{eq:PropositionTarget0}
&\lim_{\sigma\rightarrow 0}\frac{1}{\sigma}\mathbb{E}\left[\nabla_{\vect{\theta}}\vect{\pi}_{\vect{\theta}}M\left(\vect{e}-\vect c\right)\vect{e}^\top\right] \\&\qquad\qquad=\mathbb{E}\left[\nabla_{\vect{\theta}}\vect{\pi}_{\vect{\theta}}M\left(\frac{\partial\vect g}{\partial \vect d}\Sigma\frac{\partial\vect g}{\partial \vect d}^\top\right) \right]\nonumber
\end{align}
holds. Using \eqref{eq:PinvCompatible}, we finally observe that
\begin{align}
\label{eq:PropositionTarget}
&\lim_{\sigma\rightarrow 0}\frac{1}{\sigma}\mathbb{E}\left[\nabla_{\vect{\theta}}\vect{\pi}_{\vect{\theta}}M\left(\vect{e}-\vect c\right)\vect{e}^\top\right]  =  \nabla_{\vect{\theta}}\vect{\pi}_{\vect{\theta}}.
\end{align}
We can then conclude that \eqref{eq:CovarianceConditionForExactGradient} holds and that the policy gradient estimator is exact, i.e. \eqref{eq:Exact:PolicyGradient} holds.

\end{IEEEproof} 
We need to caveat here the practical implications of Proposition \ref{eq:DetPolicy:WellDefined}. First, the results hold for $\tau\rightarrow 0$, with $\sigma\rightarrow 0$. If using matrix $M$ defined via the classic inverse \eqref{eq:MeanEstimator}, the results of Proposition \ref{eq:DetPolicy:WellDefined} hold in the sense that for any $\tau$, \eqref{eq:PropositionTarget} holds asymptotically for $\sigma$ sufficiently small. Hence reducing $\tau$ may require reducing $\sigma$ for \eqref{eq:PropositionTarget} to hold. Alternatively, $M$ ought to be systematically defined via a pseudo-inverse. Unfortunately, the definition of $M$ then becomes somewhat arbitrary and non-smooth. 

Additionally, one ought to observe that the assumptions of Proposition \ref{eq:DetPolicy:WellDefined} are fairly restrictive, as they do not allow one to adjust the model or constraints in the robust MPC scheme, which leaves only the cost function as subject to adaptation. While \cite{Gros2018} shows that it is theoretically enough to adapt only the MPC cost function to generate the optimal control policy from the MPC scheme, this result requires a rich parametrization of the cost, which may be undesirable.

When the model and/or constraints of the NMPC scheme are meant to be adjusted by the RL algorithm, such that the assumptions of Proposition \ref{eq:DetPolicy:WellDefined} are not satisfied, then the policy gradient can be incorrect. The issue is associated to parameters $\vect\theta$ that can (locally) move the policy in directions orthogonal to the strictly active constraints. Indeed, for $\tau \rightarrow 0$, \eqref{eq:NLP:Disturbed}-\eqref{eq:CostGradientDisturbance} yield samples that are (for $\sigma\rightarrow 0$) in the span of $\frac{\partial \vect g}{\partial \vect d}$, which is rank deficient when $\vect\pi_{\vect\theta}$ strictly activates some constraints. However, if the assumptions of Proposition \ref{eq:DetPolicy:WellDefined} are not satisfied, $\frac{\partial \vect g}{\partial \vect \theta}$ can span directions that are in the null space of $\frac{\partial \vect g}{\partial \vect d}$, and therefore not explored. It follows that the policy gradient can be wrong in these directions. These observations are illustrated in Fig. \ref{fig:Illustration3}.

\begin{figure}
	\includegraphics[width=1\linewidth,clip,trim= 70 200 55 280]{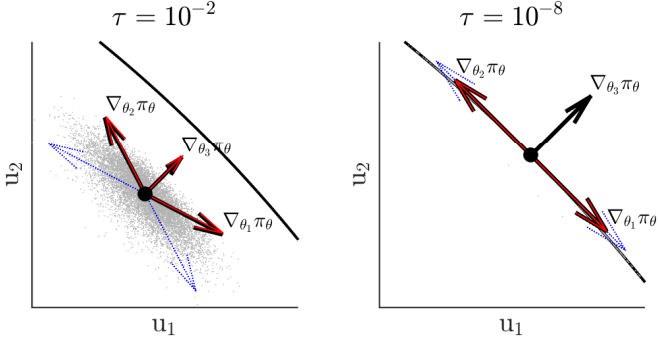} 
	\caption{\footnotesize \footnotesize Illustration of Proposition \ref{eq:DetPolicy:WellDefined} and the following discussion for the small problem \eqref{eq:TrivialExample}. The solid black arrows represent the directions spanned by $\nabla_{\vect \theta}\vect\pi_{\vect\theta}$
	. The red, dashed-line arrows report the corresponding terms in $\frac{1}{\sigma}\mathbb{E}\left[\nabla_{\vect{\theta}}\vect{\pi}_{\vect{\theta}}M\left(\vect{e}-\vect c\right)\vect{e}^\top\right]$ appearing in \eqref{eq:PropositionTarget}. The dotted-line blue arrows report the directions spanned by $\frac{\partial \vect g}{\partial \vect d}$. One can see that \eqref{eq:PropositionTarget} holds for $\tau >0$ (left graph), and holds for parameters $\vect\theta_{1,2}$ for $\tau\rightarrow 0$ (right graph) as they satisfy the assumptions of Proposition \ref{eq:DetPolicy:WellDefined}. However, \eqref{eq:PropositionTarget} does not hold for $\vect\theta_3$, as it influences the constraint \eqref{eq:TrivialExample:Const}, and therefore violates the assumptions of Proposition \ref{eq:DetPolicy:WellDefined} (right graph). One can construe the problem as a lack of exploration (blue dotted-line arrows) in the direction $\nabla_{\vect \theta_3}\vect\pi_{\vect\theta}$ due to the active constraint.}
	\label{fig:Illustration3}
\end{figure}

For cases that do not satisfy the assumptions of Proposition~\ref{eq:DetPolicy:WellDefined}, working with $\tau >0$ (although possibly small) appears to be the best option. We ought to underline here that while the corrections $M$ and $\vect c$ are in theory needed in order to build a correct policy gradient estimation \eqref{eq:Exact:PolicyGradient}, the error in the policy gradient estimation resulting from not using these corrections  is yet to be investigated in detail. 

While $\Sigma$ in \eqref{eq:d:distribution} can in principle be chosen freely, a reasonable option is to adopt $\Sigma = I$, i.e., an isotropic gradient disturbance, in which case
\begin{align}
M = \left(\frac{\partial\vect g}{\partial \vect d}\frac{\partial\vect g}{\partial \vect d}^\top\right)_{\vect d=0}^{-1}.
\end{align}
We turn now to detailing how \eqref{eq:CovMean:Estimators} can be evaluated at low computational expenses. 

\subsection{Implementation \& Sensitivity computation}

In order to compute the sensitivities required in \eqref{eq:CovMean:Estimators} to evaluate $\vect c$ and $M$, the first and second-order sensitivities of $\vect g$ are required. In turn, this requires one to evaluate the sensitivities of the relaxed KKTs \eqref{eq:NLP:Disturbed:IP}. In this section we detail how this can be done. We first observe that if LICQ and SOSC hold \cite{Nocedal2006} for the NLP \eqref{eq:NLP:Disturbed}, then $\frac{\partial \vect r_\tau}{\partial \vect z}$ is full rank, and the Implicit Function Theorem (IFT) guarantees that one can evaluate the first-order sensitivities of \eqref{eq:NLP:Disturbed:IP} by solving the linear equations:
\begin{align}
\label{eq:z:sensitivities}
\frac{\partial \vect r_\tau}{\partial \vect z}\frac{\partial \vect z}{\partial \vect d} + \frac{\partial \vect r_\tau}{\partial \vect d} = 0, \qquad
\frac{\partial \vect r_\tau}{\partial \vect z}\frac{\partial \vect z}{\partial \vect \theta} + \frac{\partial \vect r_\tau}{\partial \vect \theta} = 0,
\end{align}
for $\frac{\partial \vect z}{\partial \vect d}$ and $\frac{\partial \vect z}{\partial \vect \theta}$. 
One can then readily obtain $\frac{\partial \vect g}{\partial \vect d}$ and $\frac{\partial \vect g}{\partial \vect \theta}$  by extracting the first $n_{\vect a}$ rows of $\frac{\partial \vect z}{\partial \vect d}$, $\frac{\partial \vect z}{\partial \vect \theta}$. The Jacobian $\frac{\partial \vect g}{\partial \vect \theta}$ the delivers
\begin{align}
\nabla_{\vect\theta}\vect\pi_{\vect\theta}^\tau = \frac{\partial \vect g}{\partial \vect \theta}^\top,
\end{align}
required in \eqref{eq:DetPiGradient:Approx}, while $\frac{\partial \vect g}{\partial \vect d}$ is required in \eqref{eq:CovEstimator}.

The second-order term $\frac{\partial^2 \vect g}{\partial \vect d_i\vect d_j}$ needed in \eqref{eq:MeanEstimator} can be obtained from solving the second-order sensitivity equation of the NLP:
\begin{align}
\label{eq:SecondOrderSens:Stochastic}
&\frac{\partial \vect r_\tau}{\partial \vect z} \frac{\partial^2 \vect z}{\partial \vect d_i \partial \vect d_j} +\left(\frac{\partial^2 \vect r_\tau}{\partial \vect d_i \partial \vect z} + \sum_k\frac{\partial^2 \vect r_\tau}{\partial \vect z\partial \vect z_k}\frac{\partial \vect z_k}{\partial \vect d_i}\right)\frac{\partial \vect z}{\partial \vect d_j} +  \frac{\partial^2 \vect r_\tau}{\partial \vect d_i \partial \vect d_j}\nonumber \\& \hspace{1cm}+ \sum_k \frac{\partial^2 \vect r_\tau}{\partial \vect d_j \partial \vect z_k}\frac{\partial \vect z_k}{\partial \vect d_i}=0,
\end{align}
for $ \frac{\partial^2 \vect z}{\partial \vect d_i \partial \vect d_j}$. The sensitivity $\frac{\partial^2 \vect g}{\partial \vect d_i\vect d_j}$ is then obtained by extracting the first $n_{\vect a}$ rows of $ \frac{\partial^2 \vect z}{\partial \vect d_i \partial \vect d_j}$. Note that for computational efficiency, \eqref{eq:SecondOrderSens:Stochastic} is best treated as a tensor. We should underline here that computing the sensitivities is typically fairly inexpensive, if using an adequate algorithmic.

\section{Safe RL steps for robust linear MPC} \label{sec:RLsteps}

The methodology described so far allows one to deploy a safe policy and safe exploration using a robust NMPC scheme in order to compute the deterministic policy gradient, and determine directions in the parameter space $\vect \theta$ that improve the closed-loop performance of the NMPC scheme. However, taking a step in $\vect \theta$ can arguably jeopardize the safety of the NMPC scheme itself, e.g., by modifying the constraints, or the models underlying the scenario tree. The problem of modifying the NMPC parameters while maintaining safety is arguably a complex one, and beyond the scope of this paper. However, in this section, we propose a practical approach to handle this problem in a data-driven context. In this paper, we propose an approach readily applicable to the linear robust MPC case, see Section \ref{sec:SafeTree}.


When the dispersion set $\vect X_+\left(\vect s,\vect a\right) $ can only be inferred from data, condition \eqref{eq:DispersionApprox} arguably translates to \eqref{eq:DispersionApprox:Data}. Condition \eqref{eq:DispersionApprox:Data} translates into a condition on the admissible parameters $\vect \theta$, i.e., it specifies the parameters that are safe \textit{with respect to the data observed so far}. Condition \eqref{eq:DispersionApprox:Data} tests whether the points $\vect s_{k+1} - \vect F_0\left(\vect s_k,\vect a_k,\vect \theta\right)$ are in the polytope $\vect W$, which can be easily translated into a set of algebraic constraints imposed on $\vect\theta$. We observe that a classic gradient step of step-size $\alpha>0$ reads as:
\begin{align}
\label{eq:RL:GradientStep}
\vect\theta = \vect\theta_- - \alpha\widehat{\nabla_{\vect{\theta}}\, J(\vect{\pi}_{\vect{\theta}})} ,
\end{align}
where $\vect\theta_-$ is the previous vector of parameters. One can observe that the gradient step can be construed as the solution of the optimization problem:
\begin{align}
\min_{\vect\theta}&\quad \frac{1}{2}\left\|\vect\theta-\vect\theta_-\right\|^2+ \alpha\widehat{\nabla_{\vect{\theta}}\, J(\vect{\pi}_{\vect{\theta}})} ^\top\left(\vect\theta-\vect\theta_-\right).
\end{align}
Imposing \eqref{eq:DispersionApprox:Data} on the gradient step generating the new parameters can then be cast as the following constrained optimization problem: 
\begin{subequations}
\label{eq:Constrained:RL}
\begin{align}
\min_{\vect\theta,\vect\vartheta}&\quad \frac{1}{2}\left\|\vect\theta-\vect\theta_-\right\|^2+ \alpha\widehat{\nabla_{\vect{\theta}}\, J(\vect{\pi}_{\vect{\theta}})} ^\top\left(\vect\theta-\vect\theta_-\right) \\
\mathrm{s.t.}&\quad \vect s_{k+1} - \vect F_0\left(\vect s_k,\vect a_k,\vect \theta \right) - \sum_{i=1}^V\sum_{k=0}^{N_\mathrm{D}} \vect\vartheta_{i,k} \vect W^{i}=0, \label{eq:Inclusion1}\\
&\quad \sum_{i=1}^V \vect\vartheta_{i,k} = 1,\quad \forall k = 0,\ldots, N_{\mathcal{D}}, \\
&\quad  \vect\vartheta_{i,k} \geq 0\quad  \forall k = 0,\ldots, N_{\mathcal{D}},\quad i=1,\ldots,V, \label{eq:Inclusion3}
\end{align}
\end{subequations}
where \eqref{eq:Inclusion1}-\eqref{eq:Inclusion3} are the algebraic conditions testing \eqref{eq:DispersionApprox:Data}. We observe that unfortunately the complexity of \eqref{eq:Constrained:RL} grows with the amount of data $N_{\mathcal{D}}$ in use. In practice, the data set $\mathcal D$ should arguably be limited to incorparate relevant state transitions. A data compression technique has been proposed in \cite{Zanon2019b} to alleviate this issue in the case the nominal model $\vect F_0$ is fixed. Future work will improve on this baseline.

\section{Implementation \& illustrative Example} \label{sec:Simulations}
In this section, we provide some details on how the principle presented in this paper can be implemented, and provide an illustrative example of this implementation. At each time instant $k$, for a given state $\vect s_k$, the deterministic policy $\vect\pi_{\vect\theta}$ is computed according to \eqref{eq:NLP:Disturbed:IP} with $\vect d=0$. The solution is used to build $M$ and $\vect c$. The exploration is then generated according to \eqref{eq:NLP:Disturbed:IP} with $\vect d$ drawn from \eqref{eq:d:distribution}. We ought to underline here that, unfortunately, the NLP has to be solved twice. The data are then collected to perform the estimations \eqref{eq:A:TDEstimation} and \eqref{eq:DetPiGradient:Approx} either on-the-fly or in a batch fashion. The policy gradient estimation \eqref{eq:DetPiGradient:Approx} is then used to compute the safe parameter update according to \eqref{eq:Constrained:RL}.

\subsection{RL approach}
In the example below, a batch RL method has been used. The policy gradient was evaluated using batch Least-Squares Temporal-Difference (LSTD) techniques, whereby for each evaluation, the closed-loop system is run $S$ times for $N_t$ time steps, generating $S$ trajectory samples of duration $N_t$. The value function estimations is constructed using:
\begin{subequations}
\label{eq:LSTDV}
\begin{align}
&\sum_{k=0}^{N_t}\sum_{i= 1}^S \delta^V(\vect s_{k,i},\vect a_{k,i},\vect s_{k+1,i}) \nabla_{\vect v}\hat V^{\vect v}_{\vect{\pi}_{\vect{\theta}}}\left(\vect s_{k,i}\right) = 0  \\
&\delta^V := L(\vect s_{k,i},\vect a_{k,i})+ \gamma \hat V_{\vect{\pi}_{\vect{\theta}}}^{\vect v}\left(\vect s_{k+1,i}\right) - \hat V_{\vect{\pi}_{\vect{\theta}}}^{\vect v}\left(\vect s_{k,i}\right)
\end{align}
\end{subequations}
and based on a linear value function approximation 
\begin{align}
\hat V_{\vect{\pi}_{\vect{\theta}}}^{\vect v}\left(\vect s\right) =\vect\varrho \left(\vect s\right)^\top\vect v \label{eq:Vapprox}.
\end{align}
A simple fully parametrized quadratic function in $\vect s$ to build $\hat V^{\vect v}_{\vect{\pi}_{\vect{\theta}}}$ in the example below. Using the parameters $\vect v$ obtained from  \eqref{eq:LSTDV}, the advantage function estimation is given by:
\begin{subequations}
\label{eq:LSTDQ}
\begin{align}
&\sum_{k=0}^{N_t}\sum_{i= 1}^S \delta^Q(\vect s_{k,i},\vect a_{k,i},\vect s_{k+1,i}) \nabla_{\vect w}\hat Q_{\vect{\pi}_{\vect{\theta}}}^{\vect w}\left(\vect s_{k,i},\vect a_{k,i}\right) = 0,  \\
&\delta^Q := L(\vect s_{k,i},\vect a_{k,i})+ \gamma \hat V_{\vect{\pi}_{\vect{\theta}}}^{\vect v}\left(\vect s_{k+1,i}\right) - \hat Q_{\vect{\pi}_{\vect{\theta}}}^{\vect w}\left(\vect s_{k,i},\vect a_{k,i}\right),  \\
&\hat Q_{\vect{\pi}_{\vect{\theta}}}^{\vect w}\left(\vect s_{k,i},\vect a_{k,i}\right) = \hat V_{\vect{\pi}_{\vect{\theta}}}^{\vect v}\left(\vect s_{k,i}\right) + \hat A_{\vect{\pi}_{\vect{\theta}}}^{\vect w}\left(\vect s_{k,i},\vect a_{k,i}\right)  ,
\end{align}
\end{subequations}
where $\hat A^{\vect w}_{\vect{\pi}_{\vect{\theta}}}$ is based on \eqref{eq:Aparam:Compatible}.
We observe that both \eqref{eq:LSTDV} and \eqref{eq:LSTDQ} are linear in the parameters $\vect v$ and $\vect w$, and therefore straightforward to solve. However, they can be ill-posed on some data sets, and they ought to be solved using, e.g., a Moore-Penrose pseudo-inverse, preferably with a reasonably large saturation of the lowest singular value. The policy gradient estimation is then obtained from \eqref{eq:DetPiGradient:Approx}, using:
\begin{align}
\label{eq:DetPiGradient:Approx:FromData}
\widehat{\nabla_{\vect{\theta}}\, J(\vect{\pi}_{\vect{\theta}})} =\sum_{k=0}^{N_t}\sum_{i= 1}^S \nabla_{\vect{\theta}} \vect{\pi}_{\vect{\theta}}\left(\vect s_{k,i}\right) M\left(\vect s_{k,i}\right) \nabla_{\vect{\theta}} \vect{\pi}_{\vect{\theta}}\left(\vect s_{k,i}\right)^\top \vect w.
\end{align}

\subsection{{Robust linear MPC scheme}}
While the proposed theory is not limited to linear problems, for the sake of clarity, we propose to use a fairly simple robust linear MPC example using multiple models and process noise. We will consider the policy as delivered by the following robust MPC scheme based on multiple models and a linear feedback policy:
\begin{subequations}
\label{eq:RobustMPC:ForExamples}
\begin{align}
\min_{\vect u,\vect x}&\,\, \sum_{j=0}^{N_\mathrm{M}}\left(\left\|\vect x_{j,N} - \bar{\vect x}\right\|^2
 +\sum_{k=0}^{N-1}\left\|\matr{c}{\vect x_{j,k} - \bar{\vect x}\\ \vect u_{j,k} - \bar{\vect u}}\right\|^2\right)\\
\mathrm{s.t.}&\,\, \vect x_{j,k+1} = A_0 \vect x_{j,k} + B_0 \vect u_{j,k} + \vect b_0 +  \vect W^j, \label{eq:StateDynamics}\\
&\,\, \|\vect x_{j,k}\|^2 \leq 1,\quad \forall\, j= 0,\ldots,N_\mathrm{M},\, k= 1,\ldots N,   \label{eq:StateConst}\\
&\,\, \vect x_{j,0} = \vect s,\quad \forall \, j = 1,\ldots, N_\mathrm{M}, \\
&\,\, \vect u_{j,0} = \vect u_{k,0},\quad \forall \, k,j = 0,\ldots, N_\mathrm{M},\\
&\,\, \vect u_{j,k} = \vect u_{0,k} - K\left(\vect x_{j,k} - \vect x_{0,k}\right),\,\,\, j=1,\ldots,N_\mathrm{M},
\end{align}
\end{subequations}
where $A_0$, $B_0$, $\vect b_0$ yield the MPC nominal model corresponding to $\vect F_0$, with $\vect W^0=0$, and $\vect W^{1,\ldots M}$ capture the vertices of the dispersion set outer approximation. Hence model $j=0$ serves as nominal model and models $j=1,\ldots,N_\mathrm{M}$ capture the state dispersion over time. The linear feedback matrix $K$ is possibly part of the MPC parameters $\vect\theta$, and is a (rudimentary) structure providing a feedback $\vect\pi^\mathrm{s}$ as described in Section \ref{sec:SafeSet}. In practice, \eqref{eq:RobustMPC:ForExamples} is equivalent to a tube-based MPC.

\subsection{Simulation setup \& results}
%
%
The simulations proposed here use the same setup as the companion paper \cite{Gros2019a} treating the stochastic policy gradient case, so as to make comparisons straightforward. The experimental parameters are summarized in Tab. \ref{tab:table1} and:
\begin{align}
\vect x_{k+1} = A_\mathrm{real} \vect x_{k} + B_\mathrm{real} \vect u_{k} + \vect n,
\end{align}
where the process noise $\vect n$ is selected Normal centred, and clipped to a ball. The real system was selected as:
\begin{align}
A_\mathrm{real} &= \kappa\matr{cc}{\cos \beta & \sin \beta\\ 
\sin \beta & \cos \beta},\,  B_\mathrm{real} = \matr{cc}{1.1 & 0 \\ 0 & 0.9}.
\end{align}
The real process noise $\vect n$ is chosen normal centred of covariance $\frac{1}{3}10^{-2} I$, and restricted to a ball of radius $\frac{1}{2}10^{-2}$. The initial nominal MPC model is chosen as:
\begin{align}
A_0 &= \matr{cc}{\cos \hat\beta & \sin \hat\beta\\ 
\sin \hat\beta & \cos \hat\beta},\,  B_0= \matr{cc}{1 & 0 \\ 0 & 1},\, \vect b_0 = \matr{c}{0\\0}.
\end{align}
and $N_\mathrm{M} = 4$ with:
\begin{subequations}
\begin{align}
\vect W^1 = \frac{1}{10}\matr{c}{-1 \\ -1},\quad \vect W^2 = \frac{1}{10}\matr{c}{+1 \\ -1} \\
\vect W^3 = \frac{1}{10}\matr{c}{+1 \\ +1},\quad \vect W^4 = \frac{1}{10}\matr{c}{-1 \\ +1}.
\end{align}
\end{subequations}
\begin{table}[h!]
  \begin{center}
    \caption{\footnotesize Simulation parameters}
    \label{tab:table1}
    \begin{tabular}{l|c|c} 
   Parameter & Value & Description\\
      \hline
      $\gamma$ & 0.99 & Discount factor\\
      $\Sigma$ & $I$ & Exploration shape\\
      $\sigma$ & $10^{-3}$ & Exploration covariance\\
      $\tau$ & $10^{-2}$ & Relaxation parameter\\
      $\beta$ & $22^\circ$ & Real system parameter\\
            $\hat\beta$ & $20^\circ$ & Model parameter\\
      $N_t$ & $20$ & Sample length\\
       $S$ & $30$ & Number of sample per batch\\       
              $N$ & $10$ & MPC prediction horizon 
    \end{tabular}
  \end{center}
\end{table}

The baseline stage cost is selected as:
\begin{align}
L = \frac{1}{20}\left\|\vect x - {\vect x}_\mathrm{ref}\right\|^2 + \frac{1}{2}\left\|\vect u - {\vect u}_\mathrm{ref}\right\|^2
\end{align}
and serves as the baseline performance criterion to evaluate the closed-loop performance of the MPC scheme. 

\begin{figure}
\center
	\includegraphics[width=1\linewidth,clip,trim= 40 160 50 325]{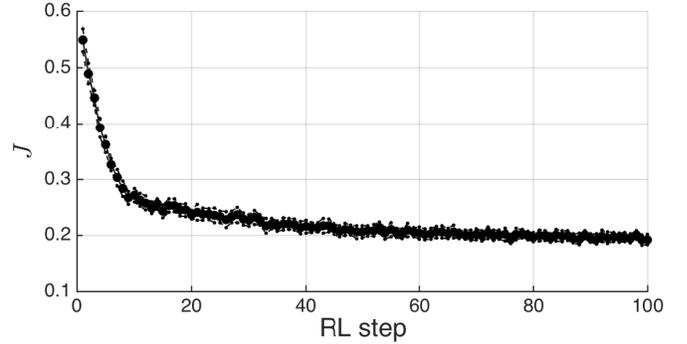} 
	\caption{\footnotesize Case 1. Evolution of the closed-loop performance $J$ over the RL steps. The solid line represents the estimation of $J$ based on the samples obtained in the batch. The dashed line represent the standard deviation due to the stochasticity of the system dynamics and policy disturbances. }
	\label{fig:J1}
\end{figure}

\begin{figure}
\center
	\includegraphics[width=0.8\linewidth,clip,trim= 40 160 40 190]{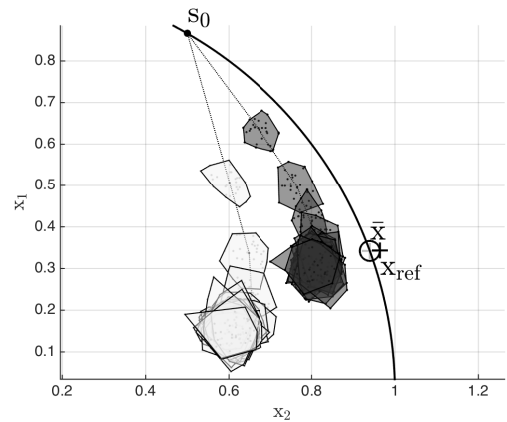} 
	\caption{\footnotesize Case 1. Closed-loop system trajectories. The initial conditions $\vect s_0$ are reported, as well as the target state reference ${\vect x}_\mathrm{ref}$ (circle), and the MPC reference $\bar{\vect x}$ at the first RL step and at the last one (grey and black $+$ symbol respectively). The trajectories at the first and last RL steps are reported as the light and dark grey polytopes. The solid black curve represents the state constraint $\|\vect x\|^2 \leq 1$.}
	\label{fig:trajX1}
\end{figure}

\begin{figure}
\center
	\includegraphics[width=1\linewidth,clip,trim= 50 180 50 390]{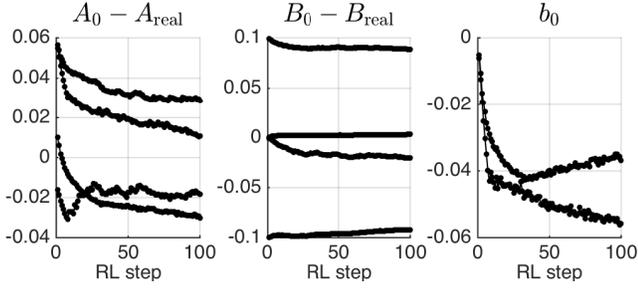} 
	\caption{\footnotesize Case 1. Evolution of the nominal MPC model over the RL steps. We report here the difference between the nominal model used in the MPC scheme and the real system.}
	\label{fig:Model1}
\end{figure}

\begin{figure}
\center
	\includegraphics[width=0.8\linewidth,clip,trim= 0 150 0 185]{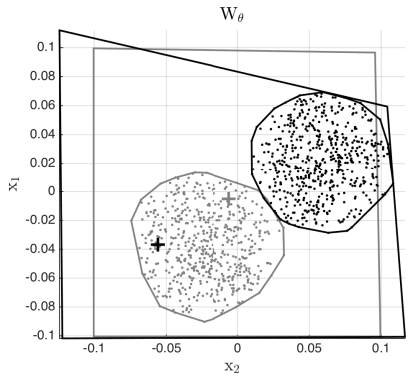} 
	\caption{\footnotesize Case 1. Evolution of the MPC model biases $\vect W^{1,\ldots M}$ over the RL steps. The light grey polytope depicts the biases at the first RL step. and the points show $\vect s_{k+1} - \vect F_0\left(\vect s_k,\vect a_k,\vect \theta\right) $ for all the samples of the first batch of data. The $+$ symbol reports the initial nominal model offset $\vect b_0$. The cloud of point is inside the grey quadrilateral thanks to the constrained RL step \eqref{eq:Constrained:RL}. The same objects are represented in black for the last step of the learning process.}
	\label{fig:Biases1}
\end{figure}

\begin{figure}
\center
	\includegraphics[width=0.8\linewidth,clip,trim= 0 150 0 185]{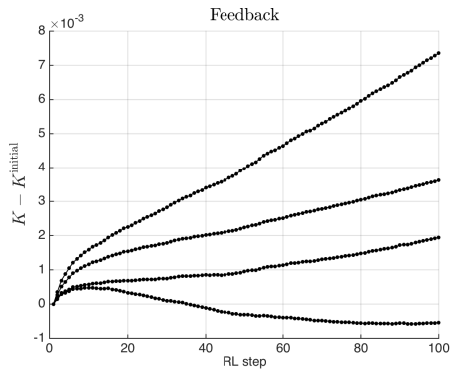} 
	\caption{\footnotesize Case 1. Evolution of the MPC feedback matrix $K$ from its initial value. The feedback is only marginally adjusted by the RL algorithm. After 100 RL steps, the adaptation of the feedback gain $K$ has not yet reached its steady-state value.}
	\label{fig:Feedback1}
\end{figure}

We considered two cases, using deterministic initial conditions $\vect s_0=\matr{cc}{\cos 60^\circ&\sin 60^\circ}^\top$. Both cases consider the parameters $\vect\theta=\left\{\bar{\vect x},\,\bar{\vect u},\, A_0,\, B_0,\,\vect b_0,\, K,\, \vect W\right\}$. The first case considers a stable real system with $\kappa = 0.95$, the second case considers an unstable real system with $\kappa = 1.05$. In both cases, the target reference $\bar{\vect x}$ was provided, together with the input reference $\bar{\vect u}$ delivering a steady-state for the nominal MPC model. The feedback matrix $K$ was chosen as the LQR controller associated to the MPC nominal model. Table \ref{tab:table1} reports the algorithmic parameters. Case 1 used a step size $\alpha= 0.05$, the second case used a step size $\alpha = 0.01$. The results for the first case are reported in Figures \ref{fig:J1}-\ref{fig:Feedback1}. One can observe in Fig. \ref{fig:J1} that the closed-loop performance is improving over the RL steps. Fig. \ref{fig:trajX1} shows that the improvement takes place via driving the closed-loop trajectories of the real system closer to the reference, without jeopardising the system safety. Fig. \ref{fig:Model1} shows how the RL algorithm uses the MPC nominal model to improve the closed-loop performance. One can readily see from Fig. \ref{fig:Model1} that RL is not simply performing system identification, as the nominal MPC model developed by the RL algorithm does not tend to the real system dynamics. 
Fig. \ref{fig:Biases1} shows how the RL algorithm reshapes the dispersion set. The upper-left corner of the set is the most critical in terms of performance, as it activates the state constraint $\|\vect x\|^2\leq 1$, and is moved inward to gain performance. The constrained RL step \eqref{eq:Constrained:RL} ensures that the RL algorithm cannot jeopardize the system safety. In Fig. \ref{fig:Feedback1}, one can see that the RL algorithm does not use much the degrees of freedom provided by adapting the MPC feedback matrix $K$.

\begin{figure}
\center
	\includegraphics[width=1\linewidth,clip,trim= 0 160 0 320]{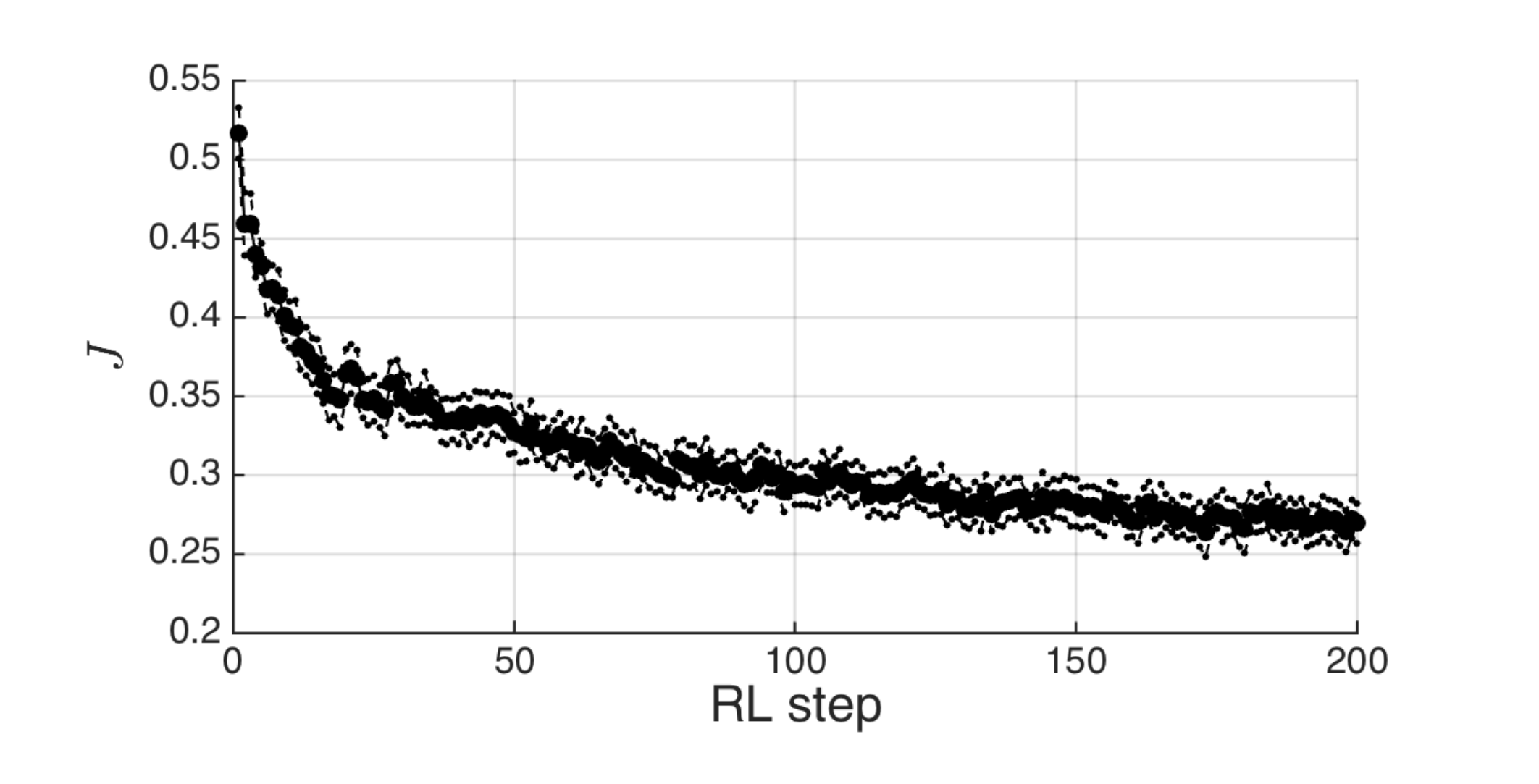} 
	\caption{\footnotesize Case 2, similar to Fig. \ref{fig:J1}.}
	\label{fig:J2}
\end{figure}
\begin{figure}
\center
	\includegraphics[width=0.8\linewidth,clip,trim= 0 150 0 180]{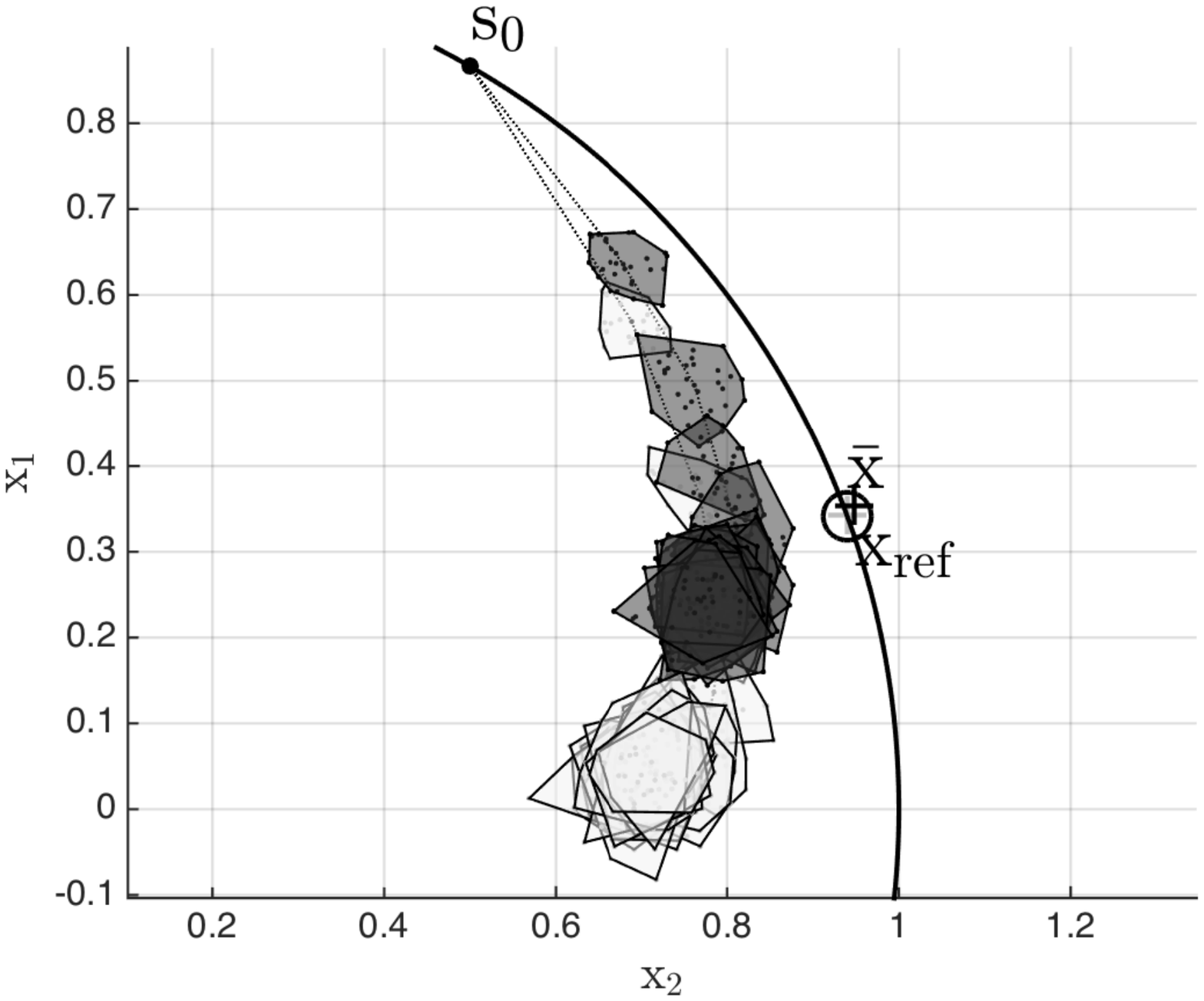} 
	\caption{\footnotesize Case 2, similar to Fig. \ref{fig:trajX1}}
	\label{fig:trajX2}
\end{figure}
\begin{figure}
\center
	\includegraphics[width=1\linewidth,clip,trim= 50 180 50 390]{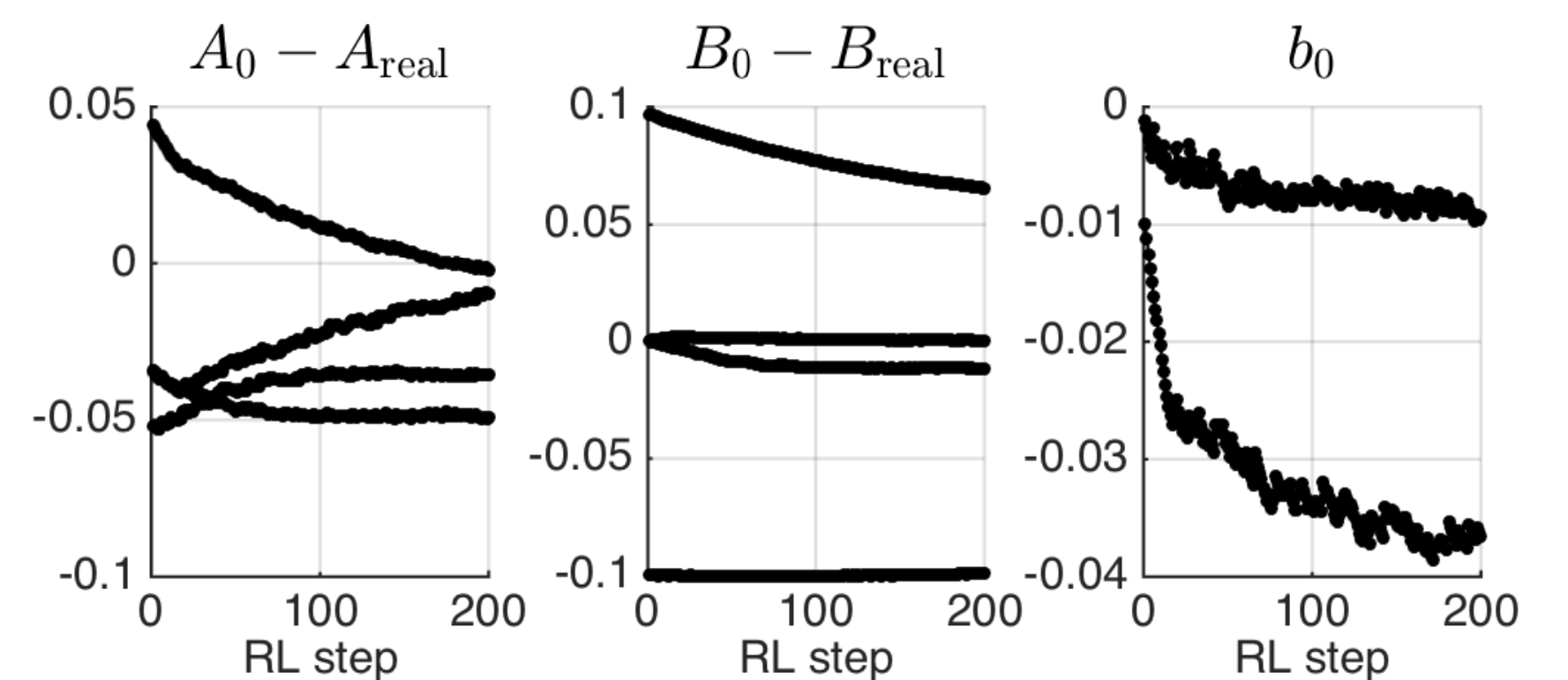} 
	\caption{\footnotesize Case 2, similar to \ref{fig:Model1}.}
	\label{fig:Model2}
\end{figure}
\begin{figure}
\center
	\includegraphics[width=0.8\linewidth,clip,trim= 0 150 0 180]{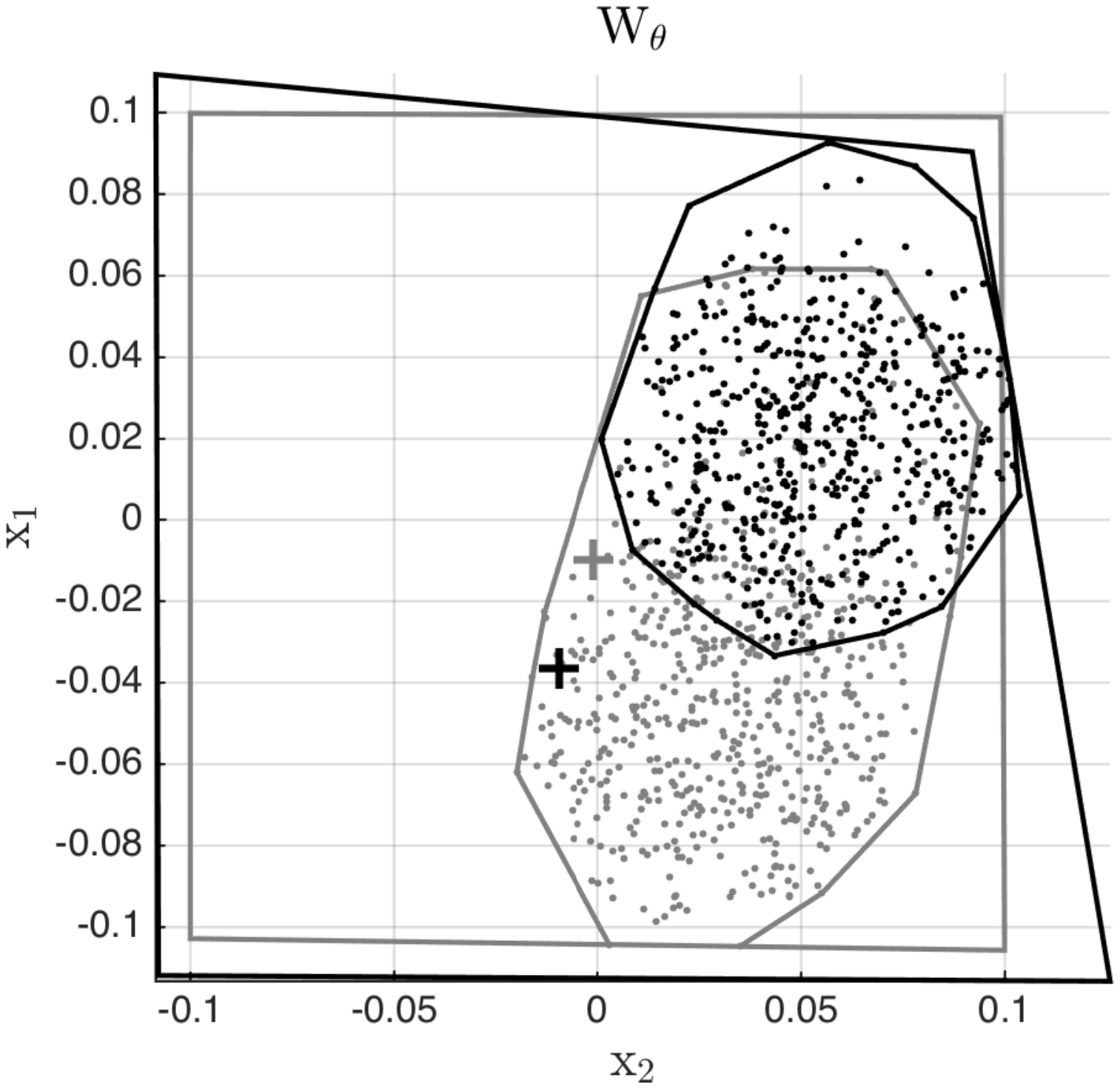} 
	\caption{\footnotesize Case 2, similar to Fig. \ref{fig:Biases1}. }
	\label{fig:Biases2}
\end{figure}
\begin{figure}
\center
	\includegraphics[width=0.8\linewidth,clip,trim= 0 150 0 185]{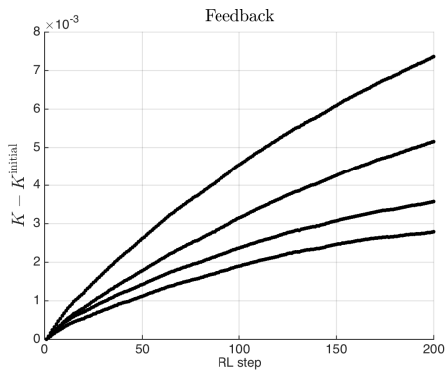} 
	\caption{\footnotesize Case 2, similar to Fig. \ref{fig:Feedback1}.}
	\label{fig:Feedback2}
\end{figure}

 The results for case 2 are reported in Figures  \ref{fig:J2}-\ref{fig:Feedback2}. Similar comments hold for case 2 as for case 1. The instability of the real system does not challenge the proposed algorithm, even though a smaller step size $\sigma$ had to be used as the RL algorithm appears to more sensitive to noise. 

\section{Conclusion}
This paper proposed a technique to deploy deterministic policy gradient methods using a constrained parametric optimization problem as a support for the optimal policy approximation. This approach allows one to impose strict safety constraints on the resulting policy. In particular, robust Nonlinear Model Predictive Control, where safety requirements can be imposed explicitly, can be selected as a parametric optimization problem. Imposing restrictions on the policy approximation creates some technical challenges when generating the exploration required to form the policy gradient. Computationally inexpensive methods are proposed here to tackle these challenges, using interior-point techniques when solving the parametric optimization problem. The specific case of robust Model Predictive Control, where the prediction model is linear, is further developed, and a methodology to impose safety requirements throughout the learning process is proposed. The proposed techniques are illustrated in simple simulations, showing their behavior. This paper has a companion paper \cite{Gros2019a} investigating the stochastic policy gradient approach in the same context as in this paper. In the simulations performed here, the stochastic policy gradient approach of \cite{Gros2019a} appears to be computationally more expensive than the approach proposed here.

\bibliographystyle{plain}
\bibliography{syscop}

\begin{IEEEbiography}[{\includegraphics[width=1in,height=1.25in,keepaspectratio]{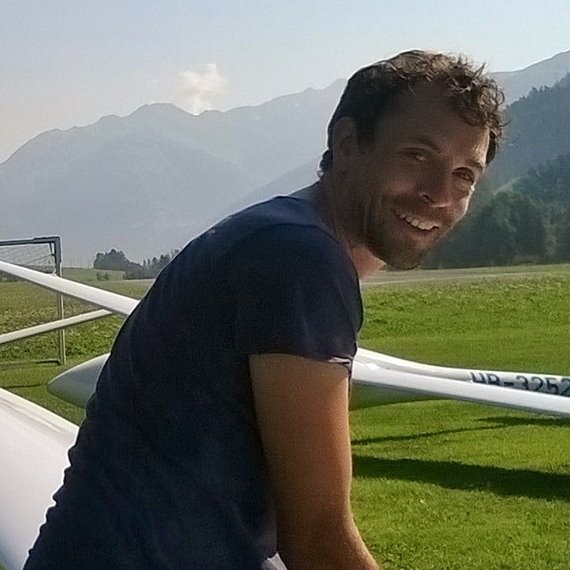}}]{S\'ebastien Gros}
	received his Ph.D degree from EPFL, Switzerland, in 2007. After a journey by bicycle from Switzerland to the Everest base camp in full autonomy, he joined a R\&D group hosted at Strathclyde University focusing on wind turbine control. In 2011, he joined the university of KU Leuven, where his main research focus was on optimal control and fast NMPC for complex mechanical systems. He joined the Department of Signals and Systems at Chalmers University of Technology, G\"{o}teborg in 2013, where he became associate Prof. in 2017. He is now full Prof. at NTNU, Norway and guest Prof. at Chalmers. His main research interests include numerical methods, real-time optimal control, reinforcement learning, and the optimal control of energy-related applications.
\end{IEEEbiography}

\begin{IEEEbiography}[{\includegraphics[width=1in,height=1.25in,clip,keepaspectratio]{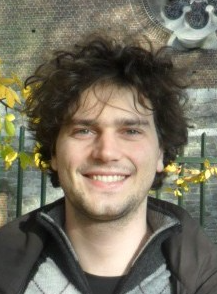}}]{Mario Zanon}
	received the Master's degree in Mechatronics from the University of Trento, and the Dipl\^{o}me d'Ing\'{e}nieur from the Ecole Centrale Paris, in 2010. After research stays at the KU Leuven, University of Bayreuth, Chalmers University, and the University of Freiburg he received the Ph.D. degree in Electrical Engineering from the KU Leuven in November 2015. He held a Post-Doc researcher position at Chalmers University until the end of 2017 and is now Assistant Professor at the IMT School for Advanced Studies Lucca. His research interests include numerical methods for optimization, economic MPC, optimal control and estimation of nonlinear dynamic systems, in particular for aerospace and automotive applications.
\end{IEEEbiography}

\end{document}